\documentclass[12pt]{article}
\usepackage[top=1in, bottom=1in, left=1.25in, right=1.25in]{geometry}
\usepackage{float}
\usepackage{color}
\usepackage{tikz}
\usepackage{setspace}
\usepackage[toc,page]{appendix}
\usepackage{amssymb}
\usepackage{showlabels}
\usepackage[normalem]{ulem}
\usepackage{multirow}
\usepackage{tikz}
\usepackage[toc]{appendix}
\usepackage{chngcntr}
\usetikzlibrary{shapes,arrows}
\usepackage{subfigure}
\usepackage[round]{natbib}
\usepackage{latexsym}
\usepackage{amsmath}
\usepackage{graphicx}
\usepackage{caption}
\providecommand{\keywords}[1]{\textbf{\textit{Key words:}} #1}
\usepackage{bm}
\usepackage[pdftex,hypertexnames=false,linktocpage=true]{hyperref}
\hypersetup{colorlinks=true,linkcolor=blue,anchorcolor=blue,citecolor=blue,filecolor=blue,urlcolor=blue,bookmarksnumbered=true,pdfview=FitB}

\newcommand{\biblist}{\begin{list}{}
		{\listparindent 0.0cm \leftmargin 0.50cm \itemindent -0.50 cm
			\labelwidth 0 cm \labelsep 0.50 cm
			\usecounter{list}}\clubpanelty4000\widowpanelty4000}
	\newcommand{\ebiblist}{\end{list}}

\newtheorem{theorem}{Theorem}
\newtheorem{lemma}[theorem]{Lemma}

\newtheorem{remark}{Remark}

\makeatother

\begin{document}
\baselineskip .3in

\title{\LARGE \bf A Profile Likelihood Approach to Semiparametric Estimation with Nonignorable Nonresponse }

\author{Hejian Sang\thanks{Google Inc., Mountain View, CA, 94043, U.S.A} \quad Kosuke Morikawa \thanks{Earthquake Research Institute, The University of Tokyo, Bunkyo-ku, Tokyo, 113-0032, Japan}}

	\maketitle
		
\begin{abstract}
Statistical inference with nonresponse is quite challenging, especially when the response mechanism is nonignorable. The existing methods often require correct model specifications for both outcome and response models. However, due to nonresponse, both models cannot be verified from data directly and model misspecification can lead to a seriously biased inference. To overcome this limitation, we develop a robust and efficient semiparametric method based on the profile likelihood. The proposed method uses the robust semiparametric response model, in which fully unspecified function of study variable is assumed. An efficient computation algorithm using fractional imputation is developed. A quasi-likelihood approach for testing ignorability is also developed.  The consistency and asymptotic normality of the proposed method are established. The finite-sample performance is examined in the extensive simulation studies and an application to the Korean Labor and Income Panel Study dataset is also presented. 
\end{abstract}


\newpage

\section{Introduction}\label{sec:intro}
Missing data is frequently encountered in statistics. The complete-case method with ignoring missing data can lead to biased estimation and misleading inference  \citep{rubin1976inference,little2014statistical}. To adjust for the bias due to missing data, some assumption about the response model is often required. If the response probability does not depend on the unobserved variable, the response mechanism is called missing at random \citep{rubin1976inference}. Otherwise, the response mechanism is called not missing at random, also referred to nonignorable missingness.  Nonignorable missingness is more challenging than missing at random, since the response model cannot be estimated from the data without extra assumptions. Furthermore, the model assumptions cannot verified from the observed data under nonignorable nonresponse.

To review the literature on nonignorable nonresponse,
let $Y$ be the study variable that is subject to missingness and $X$ be the covariate variable that is always observed. Let $\delta$ be the response indicator function of $Y$, in the sense that $\delta=1$ if $Y$ is observed, otherwise, $\delta=0$. Under the assumption of nonignorable nonresponse, \cite{diggle1994informative} propose a fully parametric method, which assumes parametric models for $f(Y|X)$ and $\mathrm{pr}(\delta=1\mid X, Y)$.The fully parametric method is very sensitive to model misspecification. \cite{scharfstein1999adjusting} , \cite{andrea2001methods} and \cite{van2012art} suggest the sensitivity analysis for the fully parametric method. Instead of assuming the parametric model for $f(Y\mid X)$, \cite{riddles2016propensity} propose an EM algorithm using fully parametric models on $f(Y\mid X, \delta=1)$. Since the data to fit  $f(Y\mid X, \delta=1)$ are fully available, the model assumption about $f(Y\mid X, \delta=1)$ can be verified from the data. However, it is still a parametric approach subject to model misspecification problem.

To achieve robustness against model misspecification, \cite{kott2010using} use a parametric model for $\mathrm{pr}(\delta=1\mid X, Y)$ and estimate the parameters by generalized method of moments. This proposed method avoids making the additional assumption on the outcome regression model. The method of \cite{kott2010using} is still subject to model misspecification of $\mathrm{pr}(\delta=1\mid X, Y)$ and is not as efficient as the maximum likelihood method.  Furthermore, \cite{morikawa2016semiparametric} propose a semiparametric maximum  likelihood method with the parametric assumption on the response model and use the nonparametric kernel method to approximate $f(Y\mid X, \delta=1)$. Note that all these methods are based on the assumption of correctly specified response model and the model specification can not be verified. To improve the robustness of the response model, \cite{kim2011semiparametric} consider a semiparametric response model. Their proposed method requires validation sample to estimate parameters in the response model. \cite{shao2016semiparametric} extend this method to avoid the requirement of validation sample. Both methods assume that response model is the generalized linear function of $Y$. Under nonignorable nonresponse, we believe that $Y$ plays a critical role in the response model. If the generalized linearity assumption of $Y$ in the response model does not hold, then the resulting estimator can be still biased.

All of these issues motivate us to propose a more robust method to handle nonignorable nonresponse. The proposed method uses the generalized partially linear model with nonparametric function of $Y$.  The estimation method is developed from the profile likelihood method. An efficient computation algorithm is proposed based on the EM algorithm using fractional imputation \citep{kim2011parametric}. Furthermore, hypothesis testing procedure is developed to test if the response mechanism is missing at random.  The proposed method is robust, since the observed regression model can be justified from the data directly and the response mechanism is an unspecified function of $Y$. 
  
The rest of this paper is organized as follows. The basic setup of nonignorable nonresponse is introduced in Section \ref{sec:setup}. The proposed method and the computation algorithm is presented in Section \ref{sec:proposal}. In Section \ref{sec:thoery}, the consistency of the proposed method and the asymptotic property are established. The performance of the proposed method is examined through simulation studies in Section \ref{sec:simulation}. The proposed method is applied to  the Korean Labor and Income Panel Study dataset in Section \ref{sec:application}. Some discussion and future work are shown in Section \ref{sec:Disscussion}. Technical proofs are given in Appendix.

\section{Setup}\label{sec:setup}
  Assume that $\{(x_1, y_1), (x_2, y_2), \cdots, (x_n, y_n)\}$ are $n$ independent and identically distributed realizations of a random vector $(X, Y)$. 
 The parameter of interest is $\theta\in \Theta$, which is uniquely determined from solving $E\left\lbrace U(\theta;X, Y) \right\rbrace=0 $. 
 Assume $x_i$ are fully observed and $y_i$ are subject to missingness. Let $\delta_i$ be the response indicator function of $y_i$, in the sense that 
 \begin{eqnarray}
 \delta_i=\left\lbrace 
 \begin{array}{ll}
 1 & \text{if $y_i$ is observed}\\
 0 & \text{otherwise}.
 \end{array}\right.\nonumber
 \end{eqnarray}
  Assume that $\{\delta_i\}_{i=1}^n$ independently follow a Bernoulli distribution with the success probability $\pi(x_i, y_i)=\mathrm{pr}(\delta_i=1| x_i, y_i)$. Then, under nonresponse, a consistent estimator of $\theta$ could be obtained by solving
  \begin{eqnarray}
  \frac{1}{n}\sum_{i=1}^{n}\frac{\delta_i}{\pi( x_i, y_i)}U(\theta; x_i, y_i)=0,\label{est2}
  \end{eqnarray}
  if the response probability $\pi(x_i, y_i)$ were known.
 
In this paper, we assume the response mechanism is not missing at random or nonignorable, in the sense that the response mechanism depends on unobserved $Y$. To estimate $\pi(x,y)$, under fully parametric assumptions, we can build the outcome model as $f(y\mid x; \zeta)$ and the response model as $\pi(x, y; \phi)$, where $(\zeta, \phi)$ are unknown parameters. Then, the observed likelihood function is 
 \begin{eqnarray}
 L_{obs}(\phi, \zeta)=\prod_{i=1}^{n}\left\lbrace \pi(x_i, y_i; \phi) f(y_i| x_i; \zeta)\right\rbrace^{\delta_i} \left[ \int \left\lbrace 1-\pi( x_i, y;\phi) \right\rbrace f(y|x_i;\zeta)dy \right]^{1-\delta_i}.\label{eq:obs_likelihood}
 \end{eqnarray}
Without additional model assumptions, maximizing $ L_{obs}(\phi, \zeta)$ in (\ref{eq:obs_likelihood}) respect to $(\phi, \zeta)$ is not identifiable.
  To avoid the non-identifiability, we also assume that 
 \begin{eqnarray}
 \mathrm{pr}(\delta_i=1\mid x_i, y_i)= \mathrm{pr}(\delta_i=1 \mid x_{i1}, y_i)=\pi( x_{i1},y_i), \nonumber
 \end{eqnarray}
 where $ x_i=(x_{i1}, x_{i2})$ and $ x_{i2}$ is the response instrumental variable \citep{wang2014instrumental}. However, the parametric assumptions cannot be verified and the fully parametric method may be sensitive to model misspecification.
 
 To achieve robustness, \cite{kim2011semiparametric} consider a semiparametric model for the response mechanism. They assume the response model can be expressed as
 \begin{eqnarray}
  \mathrm{pr}(\delta_i=1|x_i, y_i)=\frac{\exp\left\lbrace g(x_{i1})+\phi_y y_i\right\rbrace }{1+\exp\left\lbrace g( x_{i1})+\phi_y y_i\right\rbrace },\label{res}
 \end{eqnarray} 
 where $g(\cdot)$ is unspecified.   
 Note that, under assumption (\ref{res}), the predictive model for nonresponse is 
 \begin{eqnarray}
 f(y\mid  x, \delta=0)=f(y\mid  x, \delta=1)\frac{\exp(\gamma y)}{E\left\lbrace \exp(\gamma y)\mid x, \delta=1\right\rbrace   },\nonumber
 \end{eqnarray}
 where $\gamma=-\phi_y$ is the tilting parameter that describes the level of nonignorability
The consistency of the semiparametric estimation in  \cite{kim2011semiparametric} requires the correct assumption of the response model in (\ref{res}). Even though  $g(\cdot)$ is unspecified, the role of $Y$ in the response model is limited to be linear, which can be a strong assumption.

 Under the assumption of not missing at random, the function of $Y$ in the response model is very important, but can not be verifiable directly from data.  Therefore, we develop an alternative method to model the response mechanism without the generalized linearity assumption of $Y$. 
 To cover a more general class of nonignorable nonresponse, we assume the response function satisfies
\begin{eqnarray}
 \mathrm{pr}(\delta_i=1|x_i, y_i)=\frac{\exp\left\lbrace x_{i1}^T\phi+g(y_i)\right\rbrace }{1+\exp\left\lbrace x_{i1}^T \phi+g(y_i)\right\rbrace },\label{res2}
\end{eqnarray} 
 where $\phi$ is the unknown parameter and $g(\cdot)$ is an unspecified function. The proposed model in (\ref{res2}) implies that the predictive model for nonresponse is 
  \begin{eqnarray}
 f(y\mid  x, \delta=0)=f(y\mid  x, \delta=1)\frac{\exp\{-g(y)\}}{E\left[  \exp\{-g(y)\}\mid x, \delta=1\right]  }.\label{tiling2}
 \end{eqnarray}
 Hence, the proposed method can be understood as a nonparametric exponential tilting technique.
 Note that $f(y\mid x, \delta=1)$ can be estimated and validated from the observed data and $g(y)$ is unspecified.  Thus, the prediction model in (\ref{tiling2}) has less chance to suffer misspecification. 
  The details of the proposed method is presented in next Section.

 \section{Proposed method} \label{sec:proposal}
 
 Under the setup in Section \ref{sec:setup}, we assume that the semiparametric response model satisfies (\ref{res2}). Without loss of generality, we also assume that $x_{i1}$ exclude the intercept to avoid the non-identifiable issue between $x_{i1}^T\phi$ and $g(y_i)$. Denote
 \begin{eqnarray}
 \pi\left\lbrace x_{i1}^T\phi+g(y_i)\right\rbrace =\frac{\exp\left\lbrace x_{i1}^T\phi+g(y_i)\right\rbrace }{1+\exp\left\lbrace x_{i1}^T\phi+g(y_i)\right\rbrace }.\nonumber
 \end{eqnarray}
 Hence, if $g(y_i)=\phi_0+\phi_1y_i$, the proposed response model reduces to the parametric logistic model. Moreover, the proposed response mechanism degenerates to missing at random, if $g(y_i)=\phi_0$.

To estimate $\phi$ and $g(\cdot)$, the maximum profile likelihood method can be employed.  Under the complete data, the log-likelihood function can be written as
\begin{eqnarray}
&l(\phi, g)&=\sum_{i=1}^{n}\delta_i\log  \pi\left\lbrace x_{i1}^T\phi+g(y_i)\right\rbrace +(1-\delta_i)\log \left[  1- \pi\left\lbrace x_{i1}^T\phi+g(y_i)\right\rbrace \right].\label{cc}
\end{eqnarray} 
Note that, $\pi\left\lbrace x_{i1}^T\phi+g(y_i)\right\rbrace$ is a partially generalized model with nonparametric function $g$.
Then, the maximum profile likelihood method can be described as the following two steps.
\begin{itemize}
	\item []\textit{Step 1}: Fixing the parameter $\phi$, $\hat g_{\phi}(y)$ can be estimated by maximizing
	\begin{eqnarray}
	\tilde{l}_h(g\mid \phi)=\sum_{i=1}^{n}\left( \delta_i\log  \pi\left\lbrace x_{i1}^T\phi+g(y)\right\rbrace +(1-\delta_i)\log \left[  1- \pi\left\lbrace x_{i1}^T\phi+g(y)\right\rbrace \right]\right)K_h(y_i-y)\nonumber 
	\end{eqnarray}
	respect to $g(y)$, where $K_h(\cdot)$ is the kernel function with bandwidth $h$.
	\item[]\textit{Step 2}: Given the estimated function $\hat g_{\phi}(y)$, a maximum profile likelihood estimator of $\phi$ is obtained by maximizing the profile likelihood $l(\phi\mid \hat g_{\phi})$ respect to $\phi$, where
	\begin{eqnarray}
	 &l(\phi\mid \hat g_{\phi})&=\sum_{i=1}^{n}\delta_i\log  \pi\left\lbrace x_{i1}^T\phi+\hat g_{\phi}(y_i)\right\rbrace +(1-\delta_i)\log \left[ 1- \pi\left\lbrace x_{i1}^T\phi+\hat g_{\phi}(y_i)\right\rbrace \right] . \nonumber
	\end{eqnarray}
\end{itemize}
   The maximum profile likelihood estimator $\hat \phi$  converges to the asymptotic normal distribution with the rate $\sqrt{n}$. See \cite{green1985semi}, \cite{tibshirani1987local} and \cite{severini1992profile} for the estimation procedures for the generalized partial linear models.

 However, due to nonresponse, the complete log-likelihood in (\ref{cc}) is infeasible. Instead, the conditional log-likelihood, which is an unbiased estimator of  the complete log-likelihood, is used to estimate parameters under nonresponse.  The conditional log-likelihood is defined as $Q(\phi, g)= E\left\lbrace l(\phi, g)\mid \mbox{data} \right\rbrace$, which can be explicitly expressed as
 \begin{eqnarray}
Q(\phi, g)=\sum_{i=1}^{n}\left[ \delta_i\log  \pi\left\lbrace x_{i1}^T\phi+g(y_i)\right\rbrace +(1-\delta_i)E\left(  \log \left[  1- \pi\left\lbrace x_{i1}^T\phi+g(y)\right\rbrace\right] \mid x_i, \delta_i=0\right)\right]  .\label{eq:obs_like}
 \end{eqnarray} 
 Note that, in $Q(\phi, g)$,  nonresponse are integrated out by the predictive model $f(y\mid x, \delta=0)$.  The parametric model assumption about $f(y\mid x, \delta=0)$ is not justifiable due to nonresponse. Thus, we propose to use the nonparametric exponential tilting technique \citep{kim2011semiparametric} and $f(y\mid x, \delta=1)$  to avoid specifying $f(y\mid x, \delta=0)$ directly. We can rewrite  $f(y\mid x, \delta=0)$ as
 \begin{eqnarray}
 f(y\mid x, \delta=0)=f(y\mid x, \delta=1)
 \frac{\exp\left\lbrace -g(y)\right\rbrace }{E\left[ \exp\left\lbrace -g(y)\right\rbrace\mid x, \delta=1\right] },\label{t2}
 \end{eqnarray}
 where the observed outcome model $f(y\mid  x, \delta=1)$ can be validated using the observed data. Assume the parametric model for $Y$ given $ x$ and $\delta=1$ is $f(y\mid x, \delta=1; \eta)$, which is known up to $\eta$. The consistent estimator of $\eta$, say $\hat \eta$, can obtained by solving
 \begin{eqnarray}
 \sum_{i=1}^{n}\delta_i s(\eta; x_i,y_i)=0,\label{eta_s}
 \end{eqnarray}
 where $s(\eta; x_i, y_i)=\partial f(y_i\mid x_i, \delta_i=1; \eta)/\partial \eta$ is the score function of $\eta$. Using the exponential tilting technique in (\ref{t2}), $Q(\phi, g)$ in (\ref{eq:obs_like}) can be rewritten as 
 \begin{eqnarray}
 &Q(\phi, g\mid \hat \eta)&=\sum_{i=1}^{n}\delta_i\log  \pi\left\lbrace x_{i1}^T\phi+g(y_i)\right\rbrace \nonumber\\
 &&+(1-\delta_i)\frac{E\left(  \log \left[  1- \pi\left\lbrace x_{i1}^T\phi+g(y)\right\rbrace\right]\exp\left\lbrace -g(y)\right\rbrace  \mid x_i, \delta_i=1; \hat \eta\right)}{E\left[  \exp\left\lbrace -g(y)\right\rbrace  \mid x_i, \delta_i=1; \hat \eta\right] } .\nonumber
 \end{eqnarray}
 
Applying the maximum profile likelihood method to $Q(\phi, g\mid \hat \eta)$ directly is computationally intensive due to the conditional expectation. To solve this issue, we propose to apply EM algorithm using the fractional imputation method \citep{kim2011parametric}. The proposed fractional imputation algorithm is described as follows:
\begin{itemize}
	\item[] \textit{I-Step:} For the sample unit with $\delta_i=0$, generate $y_{ij}^*$ independently from $f(y\mid  x_i, \delta=1; \hat \eta)$, where $\hat\eta$ is the consistent estimator of $\eta$ from solving (\ref{eta_s}), for $j=1,2,\cdots, M$. 
	\item [] \textit{W-Step:}  Using the current value $g^{(t)}(y)$ of $g(y)$, we can assign the fractional weight as
	\begin{eqnarray}
	w_{ij}^{*(t)}\propto \exp\{- g^{(t)}(y_{ij}^*)\},\label{FI_weight}
	\end{eqnarray}
	where $\sum_j w_{ij}^*=1$. 
	
	\item[] \textit{M-Step:}  The maximum profile method can be applied to the approximation of $Q(\phi, g\mid \hat \eta)$, which is defined as
	\begin{eqnarray}
	Q(\phi, g\mid w^{*(t)}; \hat \eta)=\sum_{i=1}^{n}\left(  \delta_i \log \pi\left\lbrace x_{i1}^T\phi+g(y_i)\right\rbrace + (1-\delta_i)\sum_{j=1}^M w_{ij}^{*(t)}\log \left[1- \pi\left\lbrace x_{i1}^T\phi+g(y_{ij}^*)\right\rbrace  \right]\right) ,\nonumber
	\end{eqnarray}
	where $ w^{*(t)}$ is the set of fractional weights.
	Maximize $Q(\phi, g\mid w^{*(t)}; \hat \eta)$ using the  profile likelihood method to obtain $\phi^{(t+1)}$ and $g^{(t+1)}(\cdot)$.

\end{itemize}
Repeat \textit{W-Step} and \textit{M-Step} iteratively until the convergence is achieved.  The fractional weights in (\ref{FI_weight}) only depend on $g(\cdot)$. Since $g(\cdot)$ is modeled by a fully nonparametric function, the proposed method automatically generates the fractional weights to make 
\begin{eqnarray}
\frac{E\left(  \log \left[  1- \pi\left\lbrace x_{i1}^T\phi+g(y)\right\rbrace\right]\exp\left\lbrace -g(y)\right\rbrace  \mid x_i, \delta_i=1; \hat \eta\right)}{E\left[  \exp\left\lbrace -g(y)\right\rbrace  \mid x_i, \delta_i=1; \hat \eta\right] } \cong \sum_{j=1}^M w_{ij}^{*}\log \left[1- \pi\left\lbrace x_{i1}^T\phi+g(y_{ij}^*)\right\rbrace  \right]\nonumber
\end{eqnarray}
as close as possible. As we will show in the simulation study, the performance of the proposed method is robust to the misspecification of $x_{i1}^T\phi$ in the response model in (\ref{res2}), since the predictive model in (\ref{t2}) is free of $x_{i1}^T\phi$.
The implementation of the maximum profile likelihood method in \textit{M-Step}  is presented in the following remark.

\begin{remark}
	  The full maximization of $Q(\phi, g\mid w^*; \hat \eta)$ for each iteration of the proposed EM algorithm is not necessary.
	 \textit{M-step} can be implemented by the one-step Newton-Raphson algorithm.  Define the smoothed function of the conditional likelihood as $\tilde Q(\phi, g\mid w^{*(t)};\hat \eta)$, which can be expressed as
	\begin{eqnarray}
\sum_{i=1}^{n}\left(  \delta_i \log \pi\left\lbrace x_{i1}^T\phi+g(y)\right\rbrace K_h(y_i-y)+ (1-\delta_i)\sum_{j=1}^M w_{ij}^{*(t)}\log \left[1- \pi\left\lbrace x_{i1}^T\phi+g(y)\right\rbrace  \right]K_h(y_{ij}^*-y)\right). \label{l_tilde}
\end{eqnarray}

	 The details of \textit{M-Step} can be described as the following two steps.
	\begin{itemize}
			\item[] \textit{Step 1:} We can update $\phi$ by 
		\begin{eqnarray}
		\phi^{(t+1)}=\phi^{(t)}-B_t^{-1}A_t,\nonumber
		\end{eqnarray}
		where 
		\begin{eqnarray}
		A_t=\left.\bigtriangledown Q(\phi, \hat g_{\phi}\mid w^{*(t)}; \hat\eta)\right|_{\phi=\phi^{(t)}}\nonumber
		\end{eqnarray}
		is the marginal gradient,  $ Q(\phi, \hat g_{\phi}\mid w^{*(t)}; \hat\eta)$ is the profiled function of $\phi$, and
		\begin{eqnarray}
		B_t=\left.\bigtriangleup Q(\phi, \hat g_{\phi}\mid w^{*(t)})\right|_{\phi=\phi^{(t)}}\nonumber
		\end{eqnarray}
		is the Hessian matrix.

		\item []\textit{Step 2:} Update $g(y)$ by 
		\begin{eqnarray}
		g^{(t+1)}(y)= g^{(t)}(y)-\frac{G_t(y)}{H_t(y)},\nonumber
		\end{eqnarray}
		where
		\begin{eqnarray}
	     G_t(y)=\left.\bigtriangledown \tilde Q(\phi, g(y)\mid w^{*(t)}; \hat \eta)\right|_{\phi=\phi^{(t+1)}, g=g^{(t)}}\nonumber
		\end{eqnarray}
		is a gradient of the smoothed function $\tilde Q(\phi, g(y)\mid w^{*(t)}; \hat \eta)$ in (\ref{l_tilde}) respect to $g(y)$
		 and 
		\begin{eqnarray}
	    H_t(y)=\left.\bigtriangleup \tilde Q(\phi, g(y)\mid w^{*(t)};\hat \eta)\right|_{\phi=\phi^{(t+1)}, g=g^{(t)}}\nonumber
		\end{eqnarray}
		is a Hessian of $\tilde Q(\phi, g(y)\mid w^{*(t)};\hat \eta)$ respect to $g(y)$
	
	\end{itemize}
	
	The technical derivations in the  \textit{Step 1} and \textit{Step 2} are shown in Appendix \ref{App:App_A}.
	
\end{remark}

Once the convergence of the proposed EM algorithm is achieved, the final estimator of $\theta$, say $\hat \theta$, can be obtained by solving 
\begin{eqnarray}
\frac{1}{n}\sum_{i=1}^{n}\frac{\delta_i}{\pi\left\lbrace  x_{i1}^T\hat{\phi}+\hat g(y_i)\right\rbrace } U(\theta; x_i, y_i)=0.\label{est}
\end{eqnarray}

\begin{remark}
	Alternatively, we can also estimate $\theta$ by solving
	\begin{eqnarray}
	\sum_{i=1}^{n}\left\lbrace \delta_i U(\theta;x_i, y_i)+(1-\delta_i)\sum_{j=1}^M w_{ij}^*U(\theta; x_i, y_{ij}^*)\right\rbrace =0,\nonumber
	\end{eqnarray}
	which is an empirical approximation of 
	\begin{eqnarray}
	\sum_{i=1}^n \left[ \delta_i U(\theta;x_i, y_i)+(1-\delta_i) E\left\lbrace U(\theta; x_i, y)\mid x_i, \delta_i=0 \right\rbrace \right] =0.\nonumber
	\end{eqnarray}
	In this paper, we focus on the estimator in (\ref{est}).
\end{remark}

\begin{remark}
	Note that, if $Y$ is binary, then the proposed method is degenerated to the parametric model.  The response mechanism is 
	\begin{eqnarray}
	\mathrm{pr}(\delta=1\mid x, y)=\frac{ \exp\left\lbrace x_{1}^T\phi+g(y)\right\rbrace}{1+\exp\left\lbrace x_{1}^T\phi+g(y)\right\rbrace},
	\end{eqnarray}
	which is a parametric function of $\left\lbrace \phi, g(0), g(1)\right\rbrace $.
	For a general discrete $Y$, the proposed method still works by employing the  kernel smoothing for discrete variables in \cite{hall1981nonparametric} and \cite{chen2011nonparametric}.
\end{remark}

\begin{remark}
	It is worth to mentioning that the parametric observed regression model $f(y\mid x, \delta=1; \eta)$ can be replaced by a nonparametric regression model.  We can show that for the function $A(\delta,  x_1, Y)=\log\left\lbrace 1-\pi(\phi, g; x_1, Y)\right\rbrace $, we can express 
	\begin{eqnarray}
	E\left\lbrace A(\delta, x_1, Y)\mid x, \delta=0\right\rbrace =\frac{\int A(\delta,  x_1, y) O(\phi, g; x_1, y)f(y\mid x, \delta=1)dy}{\int O(\phi, g; x_1, y)f(y\mid x, \delta=1)dy},\nonumber
	\end{eqnarray}
	where
	\begin{eqnarray}
	O(\phi, g;x_{i1},y_i)=\frac{\mathrm{pr}(\delta=0\mid x_i, y_i)}{\mathrm{pr}(\delta=1\mid x_i, y_i)},\nonumber
	\end{eqnarray}
	which leads to $O(\phi, g;x_{i1},y_i)=\exp\left\lbrace - x_{i1}^T\phi-g(y_i)\right\rbrace$ under the model assumption in (\ref{res2}).
	Thus, using the kernel smoothing method, we can approximate $E\left\lbrace A(\delta, x_1, Y)\mid x, \delta=0\right\rbrace$ as 
	\begin{eqnarray}
	\hat E\left\lbrace A(\delta, x_1, Y)\mid x, \delta=0\right\rbrace=\frac{\sum_{j=1}^{n}\delta_j K_{H}( x_j- x)O(\phi,g;  x_{1},y_j)A(\delta, x_1, y_j)}{\sum_{j=1}^{n}\delta_j K_{H}( x_j- x)O(\phi,g; x_{1},y_j)},\label{ae}
	\end{eqnarray}
	where $K(\cdot)$ is the kernel function and $H$ is a diagonal bandwidth matrix.
	Since we have already shown that $O(\phi,g; x_1, y)=\exp\left\lbrace-\phi^T x_{1}-g(y) \right\rbrace $, we can simply (\ref{ae}) as
	\begin{eqnarray}
	\hat E\left\lbrace A(\delta, x_1, Y)\mid x, \delta=0\right\rbrace=\frac{\sum_{j=1}^{n}\delta_j K_{H}( x_j- x)\exp\left\lbrace -g(y_j)\right\rbrace A(\delta, x_1, y_j)}{\sum_{j=1}^{n}\delta_j K_{H}( x_j- x)\exp\left\lbrace -g(y_j)\right\rbrace}.\label{ce}
	\end{eqnarray}
	Using (\ref{ce}) to replace the conditional expectation in $Q(\phi, g\mid \hat \eta)$, we can build the conditional log-likelihood function without any parametric assumption for the observed outcome model $f(y\mid x, \delta=1)$.
\end{remark}

\section{Asymptotic Theory}\label{sec:thoery}

In this section, we establish  consistency and asymptotic normality of the proposed estimator in (\ref{est}).  We summarize the sufficient conditions for asymptotic theories as follows. The assumptions in details are presented in Appendix \ref{App:App_C}.
\begin{itemize}
	\item[(\textbf{C1}):] The true response model $\pi(x, y)$ satisfies (\ref{res2}).
	\item[(\textbf{C2}):] The kernel function $K(\cdot)$ satisfies the following properties
	\begin{itemize}
		\item [] $K(u)=0$ for $|u|>1$;
		\item [] $\sup_u|K(u)|<\infty$;
		\item[] $\int K(u)du=1, \int uK(u)du=0, \int u^2K(u)< \infty$.
	\end{itemize}
\item[(\textbf{C3}):] Regularity conditions to establish the asymptotic normality of $\hat \eta$.
\item [(\textbf{C4}):] Regularity conditions for the partially logistic linear models.

\item [(\textbf{C5}):] Regularity conditions for the estimating equation $ U(\theta; X,Y)$.
\end{itemize}

Condition (\textbf{C1}) is our semiparametric model assumption. (\textbf{C2}) is a standard assumption for the kernel regression method. The regularity conditions in (\textbf{C3}) are standard to obtain the asymptotic normality of maximum likelihood estimator $\hat \eta$. (\textbf{C4}) introduces the sufficient conditions to establish the asymptotic normality of $\hat \phi$ under the complete data. (\textbf{C5}) are the regularity conditions for the  estimating equation. The details of (\textbf{C3} )to (\textbf{C5}) are shown in Appendix \ref{App:App_C}.
\begin{lemma}\label{Monotone}
	Under Conditions (\textbf{C1})--(\textbf{C4}), our proposed algorithm enjoys the monotone increasing property, in the sense of 
	\begin{eqnarray}
		Q(\phi^{(t)}, g_{\phi^{(t)}}\mid w^{*(t)}; \hat \eta)\leq Q(\phi^{(t+1)}, g_{\phi^{(t+1)}}\mid w^{*(t)}; \eta)\label{m_phi},\\
	\tilde Q(\phi^{(t+1)}, g^{(t)}\mid w^{*(t)};\hat \eta)\leq\tilde Q(\phi^{(t+1)}, g^{(t+1)}\mid w^{*(t)}; \hat \eta),\label{smooth_g}
	\end{eqnarray}
	where $\tilde{Q}(\phi, g\mid w^{*(t)};\hat \eta)$ is defined in (\ref{l_tilde})
	for any $y$.
\end{lemma}

The proof of Lemma \ref{Monotone} is shown in Appendix \ref{App:App_C}.
From Lemma (\ref{Monotone}), the estimators from our proposed EM algorithm lead to  the monotone increase of the profiled conditional log-likelihood of $\phi$ and the smoothed conditional log-likelihood of $g$. 

\begin{theorem}\label{thm:thm1}
	Under conditions (\textbf{C1})--(\textbf{C4}), we have
	\begin{eqnarray}
	\sqrt n (\hat \phi -\phi_{0})\xrightarrow{}N(0, \Sigma_0),
	\end{eqnarray}
	in distribution, as $n,M\xrightarrow{}\infty$. $\phi_{0}$ is the true parameter value and $\Sigma_0=\Sigma_1+\Sigma_2+\Sigma_3$.
	$\Sigma_1$ is the observed Fisher information. $\Sigma_2$ is the variability of estimating $\eta_0$ and $\Sigma_3$ is the covariance between $\hat \phi$ and $\hat \eta$.
\end{theorem}
The proof of Theorem \ref{thm:thm1} is presented in Appendix \ref{App:App_C}.
From Theorem \ref{thm:thm1}, we can see that our proposed method has the $\sqrt{n}$ convergence rate for parameters, which is the same for fully parametric models.

\begin{theorem}\label{thm:thm2}
	Under conditions (\textbf{C1})--(\textbf{C5}), we can establish that 
	\begin{eqnarray}
	\sqrt n (\hat \theta-\theta_0)\xrightarrow{L}N(0, \Sigma),
	\end{eqnarray}
	where $\theta_0$ is the true value and $\Sigma>0$.
\end{theorem}
The proof of Theorem \ref{thm:thm2} is shown in Appendix \ref{App::D}. In Appendix \ref{App::D}, we have
\begin{eqnarray}
\hat \theta-\theta_0\cong-\left[ E\left\lbrace  \frac{\partial U(\theta_0\mid \phi_0, g_0)}{\partial \theta_0} \right\rbrace \right] ^{-1} \left[  U(\theta_0\mid \phi_0, g_0) +E\left\lbrace \frac{\partial U(\theta_0\mid \phi_0,  g_0)}{\partial \phi_0}\right\rbrace (\hat \phi-\phi_0) \right] .\nonumber
\end{eqnarray}
Then, we can see that $\Sigma$ is a composited variability from estimating equation in (\ref{est}) and the profiled function of $\phi_0$ in $Q(\phi, g\mid \eta_0)$.

\section{Ignorability Test}\label{sec:test}
In Section \ref{sec:setup}, we assume the response mechanism satisfies (\ref{res2}). Hence, if $g(y)$ is a constant, say $g(y)=c$ for some $c\in \mathbb{R}$, the response mechanism degenerates to missing at random. If we are confident that the response mechanism is missing at random,  estimation and inference can be greatly simplified without worrying about nonignorable nonresponse bias.
Since our response model is a nonparametric model of $Y$, it is a great interest to test if the response mechanism is missing at random without specifying $g$.

Under the null hypothesis $\mathrm{H}_0: g(y)=c$, the response mechanism is a parametric model of unknown $(\phi,c)$. Furthermore, $(\phi,c)$ can be estimated from maximizing the log-likelihood function of $(\phi,c)$. Specifically, that is to maximize
\begin{eqnarray}
l(\phi,c)=\sum_{i=1}^{n}\delta_i\log \pi(\phi, c; x_i)+(1-\delta_i)\log\left\lbrace 1-\pi(\phi, c; x_i) \right\rbrace, \label{like_mar}
\end{eqnarray}
respect to $(\phi, c)$, where
\begin{eqnarray}
\pi(\phi, c; x_i)=\frac{\exp(x_{i1}^T\phi+c)}{1+\exp(x_{i1}^T\phi+c)}.\nonumber
\end{eqnarray}
 Note that, the likelihood ratio test statistic can not be used here due to the non-negligible smoothing bias and different likelihood functions (smoothed and unsmoothed functions). See \cite{hardle1998testing} and \cite{lombardia2008semiparametric} for related clarification.
To solve this issue, \cite{hardle1998testing} proposed using the weighted distance test statistic based on the quasi-likelihood of the logistic model. Under complete response, we propose using
\begin{eqnarray}
R=\sum_{i=1}^{n}\pi(\hat\phi_a,\hat c_a; x_i)\left\lbrace 1-\pi(\hat \phi_a,\hat c_a; x_i) \right\rbrace\left\lbrace x_{i1}^T(\hat \phi-\hat \phi_a)+\hat g(y_i)-\hat c_a\right\rbrace^2, \label{test1}  
\end{eqnarray}
where $(\hat \phi_a, \hat c_a)$ is the solution of (\ref{like_mar}) and $\hat \phi$ is the estimator of the proposed profile method. Under the null hypothesis and some regularity conditions, \cite{hardle1998testing} show
\begin{eqnarray}
v_n^{-1}(R-e_n)\xrightarrow{}N(0,1),\nonumber
\end{eqnarray}
in distribution. However, $(v_n,e_n)$ are very difficult to compute.
Under nonresponse, the test statistic in (\ref{test1}) can be approximated by 
\begin{eqnarray}
&\hat R=&\sum_{i=1}^{n}\pi(\hat\phi_a,\hat c_a; x_i)\left\lbrace 1-\pi(\hat \phi_a,\hat c_a; x_i) \right\rbrace\left[ \delta_i\left\lbrace x_{i1}^T(\hat \phi-\hat \phi_a)+\hat g(y_i)-\hat c_a\right\rbrace^2\right.\nonumber\\
&&\left. +(1-\delta_i)\sum_{j=1}^{M}w_{ij}^*\left\lbrace x_{i1}^T(\hat \phi-\hat \phi_a)+\hat g(y_{ij}^*)-\hat c_a\right\rbrace^2\right]. \label{test2} 
\end{eqnarray}

\begin{remark}
	Note that, under the null hypothesis, 
	\begin{eqnarray}
	\sum_{j=1}^{M}w_{ij}^*\left\lbrace x_{i1}^T(\hat \phi-\hat \phi_a)+\hat g(y_{ij}^*)-\hat c_a\right\rbrace^2-\left[  x_{i1}^T(\hat \phi-\hat \phi_a)+E\left\lbrace \hat g(y)\mid \hat \eta, \delta=1,x_i\right\rbrace -\hat c_a\right]^2\xrightarrow{}0,\nonumber
	\end{eqnarray}
	almost surely, as $M\xrightarrow{}\infty$. Thus, we can rewrite 
	\begin{eqnarray}
	\hat R=R+\sum_{i=1}^{n}\pi(\hat\phi_a,\hat c_a; x_i)\left\lbrace 1-\pi(\hat \phi_a,\hat c_a; x_i) \right\rbrace(1-\delta_i)\left[\hat g(y_i)- E\left\lbrace \hat g(y)\mid \hat \eta, \delta=1,x_i\right\rbrace\right]^2.\nonumber
	\end{eqnarray}
	Under the null hypothesis, $E\left[\hat g(y_i)- E\left\lbrace \hat g(y)\mid \hat \eta, \delta=1,x_i\right\rbrace\right]^2=o_p(1)$. Thus, 
	$\hat R=R\left\lbrace 1+o_p(1)\right\rbrace $. We can conclude that $v_n^{-1}(\hat R-e_n)$ also converges to the normal distribution. If $M$ is finite, $v_n$ can be inflated by the variability of imputation and $\hat \eta$.
\end{remark}
Since $(v_n, e_n)$ is difficult to compute, and the uncertainty of imputation needs to be incorporated properly, we propose to use the bootstrap method to test $\mathrm{H_a}: g(y)=c$. Under $\mathrm{H}_0: g(y)=c$, the parametric bootstrap is developed. The algorithm of the parametric bootstrap is shown in Appendix \ref{App:App_B}.

\section{Simulation Study}\label{sec:simulation}
\subsection{Simulation Study I}
In this simulation study, we investigate the performance of the proposed method in the finite sample.  The robustness of the proposed method is also examined when the model assumption is violated.  The simulation study can be described as a $3\times 9$ factorial design, where the factors are the outcome regression model and the response mechanism. Assume the covariate $ x_i=(x_{i1}, x_{i2})$ are generated from $N( u, \Sigma)$ with $ u=(1,1)^T$ and $\Sigma=\text{Diag}(0.25,0.25)$ independently.  For the outcome regression model, let $y_i=m( x_i)+e_i$, where the mean function $m(x)$ is one of followings:
\begin{eqnarray}
\mathcal M_1: & m(x)=-1+(x_2-0.5)^2\nonumber\\
\mathcal M_2: & m(x)=-2.75+x_1+x_2+x_1x_2\nonumber\\
\mathcal M_3: & m(x)=-1.75+x_1+x_2\nonumber
\end{eqnarray}
and $e_i\sim N(0, 0.25)$ independently.

For the response mechanism, let $\delta_i$ be generated from a Bernoulli distribution with the success probability $\pi_i$ independently. 
For the true response mechanism, we consider follows:

\begin{itemize}
	\item [$\mathcal R_1$:] (Linear MAR)
	\begin{eqnarray}
	\pi_i=\frac{\exp(\phi_0+\phi_1x_{i1})}{1+\exp(\phi_0+\phi_1x_{i1})},\nonumber
	\end{eqnarray}
	where $(\phi_0, \phi_1)=(0.7,0.2)$. 
	\item [$\mathcal R_2$:] (Linear NMAR)
	\begin{eqnarray}
	\pi_i=\frac{\exp(\phi_0+\phi_1x_{i1}+\phi_2y_i)}{1+\exp(\phi_0+\phi_1x_{i1}+\phi_2y_i)}, \nonumber
	\end{eqnarray}
	where $(\phi_0, \phi_1)=(1,0.2,0.2)$.
	\item [$\mathcal R_3$:] (Non-linear NMAR with quadratic term in $y$)
	\begin{eqnarray}
	\pi_i=\frac{\exp\left(\phi_0+\phi_1x_{i1}+\phi_2y_i^2 \right) }{1+\exp\left(\phi_0+\phi_1x_{i1}+\phi_2y_i^2 \right)}, \nonumber
	\end{eqnarray}
	where $(\phi_0,\phi_2, \phi_2)=(0,0.1,0.7)$.
	\item[ $\mathcal R_4$:] (Non-linear NMAR with quadratic term in both $x$ and $y$)
	\begin{eqnarray}
		\pi_i=\frac{\exp\left\lbrace \phi_0+\phi_1x_{i1}^2+\phi_2y_i^2\right\rbrace  }{1+\exp\left\lbrace \phi_0+\phi_1x_{i1}^2+\phi_2y_i^2 \right\rbrace }, \nonumber
	\end{eqnarray}
	where $(\phi_0,\phi_1,\phi_2)=(0,0.1,0.5)$.
	\item [$\mathcal R_5$:] (Non-linear NMAR with exponential term in $x_1$ and quadratic term in $y$)
	\begin{eqnarray}
	\pi_i=\frac{\exp\left\lbrace \phi_0+\phi_1\exp(x_{i1}-1)+\phi_2y_i^2\right\rbrace }{1+\exp\left\lbrace \phi_0+\phi_1\exp(x_{i1}-1)+\phi_2y_i^2\right\rbrace}, \nonumber
	\end{eqnarray}
	where $(\phi_0,\phi_1,\phi_2)=(0,0.1,0.6)$
	\item[ $\mathcal R_6$:] (Non-linear NMAR with exponential term in $y$ and interaction term)
	\begin{eqnarray}
	\pi_i=\frac{\exp\left\lbrace \phi_0+\phi_1x_{i1}y_i+\phi_2y_i^2\right\rbrace  }{1+\exp\left\lbrace \phi_0+\phi_1x_{i1}y_i+\phi_2y_i^2 \right\rbrace }, \nonumber
	\end{eqnarray}
	where $(\phi_0, \phi_1, \phi_2)=(0,0.1,0.6)$.
	
	\item[$\mathcal R_7$:](Probit NMAR)\begin{eqnarray}
	\pi_i=\Phi(\phi_0+\phi_1x_{i1}+\phi_2y_i^2),\nonumber
	\end{eqnarray}
	where $(\phi_0,\phi_1, \phi_2)=(0,-0.1,0.6)$ and $\Phi(\cdot)$ is the normal cumulative distribution function.
	
	\item[$\mathcal R_8$:](Complementary log-log NMAR)\begin{eqnarray}
	\pi_i=1-\exp\left\lbrace -\exp(\phi_0+\phi_1x_{i1}+\phi_2y_i^2)\right\rbrace ,\nonumber
	\end{eqnarray}
		where $(\phi_0,\phi_1, \phi_2)=(0,-0.05,0.3)$.
	\item[$\mathcal R_9$:]($x_1$ instrumental variable)\begin{eqnarray}
		\pi_i=\frac{\exp\left(\phi_0+\phi_1x_{i2}+\phi_2y_i^2 \right) }{1+\exp\left(\phi_0+\phi_1x_{i2}+\phi_2y_i^2 \right)}, \nonumber
		\end{eqnarray}
		where $(\phi_0,\phi_1, \phi_2)=(0,0.1,0.7)$.
\end{itemize}
 
 The response mechanism $\mathcal{R}_1$ is missing at random, in the sense of $g(y)=\phi_0$. $\mathcal{R}_2$ is the logistic linear model assumption, which is mostly used to fit the nonresponse model in \cite{kim2011semiparametric} and \cite{shao2016semiparametric}. $\mathcal R_3$ satisfies all model assumptions of the proposed method. $\mathcal R_4$ and $\mathcal{R}_5$ violate the linearity assumption of $x_{i1}$ and $\mathcal R_6$ has the interaction term of $x_i, y_i$, which leads to failure of the linearity assumption. $\mathcal R_7$ and $\mathcal R_8$ are used to check the robustness of the link function. $\mathcal{R}_9$ is used to check the violation of the instrumental variable assumption.

For each response mechanism, the overall response rates are approximately 70\%.
For each setup, we generate a Monte Carlo sample with $n=500$  independently for replication $B=2,000$. Suppose we are interested in $\theta=E(y)$. Thus, $U(\theta; x, y)=y-\theta$. For each realized sample, we apply the following methods.
\begin{itemize}
	\item [1.] Full estimator $\theta_{full}$: Use the full sample to estimate $\theta$, but which is not practical in real data analysis.
	\item [2.] CC estimator $\theta_{CC}$: Ignore nonresponse and only use responses to estimate $\theta$.
	\item[3.]  \citet{kott2010using}'s method $\theta_{KC}$: Assume the response model is 
	\begin{eqnarray}
	Pr(\delta_i=1\mid x_i, y_i)=\pi(\phi;y_i)=\frac{\exp(\phi_0+\phi_1x_{1i}+\phi_2y_i)}{1+\exp(\phi_0+\phi_1x_{1i}+\phi_2y_i)}.\label{eq:res_asumme}
	\end{eqnarray}
	The estimator can be obtained by solving 
	\begin{eqnarray}
	\frac{1}{n}\sum_{i=1}^{n}\left\lbrace\frac{\delta_i}{\pi(\phi;x_{1i},y_i)}-1 \right\rbrace (1, \bm x_i)'=\bm 0,\nonumber\\
	\frac{1}{n}\sum_{i=1}^{n}\frac{\delta_i}{\pi(\phi;x_{1i},y_i)}(y_i-\theta)=0.\nonumber
	\end{eqnarray}
	\item[4.] \cite{riddles2016propensity}'s method $\theta_{FI}$: The observed regression model is 
	\begin{eqnarray}
	y_i\mid x_i, \delta_i=1\sim N(\beta_0+\beta_1x_{i1}+\beta_2x_{i2}+\beta_3x_{i1}^2+\beta_4x_{i2}^2+\beta_5x_{i1}x_{i2}, \sigma^2).
	\end{eqnarray}
	The response working model uses (\ref{eq:res_asumme}).
	
	\item [5.] $\theta_{\text{SP}}$: The proposed method with $x_2$ as the response instrumental variable. The bandwidths are chosen by rule of thumb \citep{silverman1986density}. The working observed regression model is specified as $y_i\mid  x_i, \delta_i=1\sim N(\beta_0+\beta_1x_{i1}+\beta_2x_{i2}+\beta_3x_{i1}^2+\beta_4x_{i2}^2+\beta_5x_{i1}x_{i2}, \sigma^2)$.
	
\end{itemize}
	
The simulation results for $\mathcal R_1$ -- $\mathcal R_3$, $\mathcal R_4$ -- $\mathcal R_6$ and $\mathcal R_7$ -- $\mathcal R_9$ are presented in Table \ref{tbl1}, \ref{tbl2} and \ref{tbl3}, separately.

\begin{table}
	\centering
	\caption{Simulation results (part I) from $B=2,000$ Monte Carlo studies}\label{tbl1}
	\begin{tabular}{lllrrrrr}
		\hline
		Res & Model & Estimates & $\theta_{full}$ & $\theta_{CC}$ & $\theta_{KC}$  & $\theta_{FI}$ & $\theta_{SP}$\\
		\hline 
		\multirow{6}{*}{$R_1$} & \multirow{3}{*}{$M_1$} & bias & -0.001 & -0.002 & -0.003 & -0.002 & -0.005\\
		& & std & 0.035& 0.042 & 0.045 & 0.041 & 0.039 \\
		& & rmse & 0.035 & 0.042 & 0.045& 0.041 & 0.039\\
		& \multirow{3}{*}{$M_2$} & bias & 0.001& 0.030 & 0.001 & 0.001& -0.000\\
		& & std & 0.067 & 0.080 & 0.070 & 0.069 & 0.070 \\
		& & rmse & 0.067 & 0.085 & 0.070 & 0.069 & 0.070\\
		& \multirow{3}{*}{$M_3$} & bias & 0.000& 0.015 & 0.000 & 0.000& 0.000\\
		& & std & 0.038 & 0.045 & 0.044 & 0.044 & 0.042 \\
		& & rmse & 0.038& 0.048 & 0.044 & 0.044 & 0.042\\
		\hline
		\multirow{6}{*}{$R_2$} & \multirow{3}{*}{$M_1$} & bias & 0.001 & 0.027 & -0.000 & -0.000 & 0.003\\
		& & std & 0.035 & 0.041 & 0.043 & 0.039 & 0.039 \\
		& & rmse & 0.035 & 0.049 & 0.043 & 0.039 & 0.039 \\
		& \multirow{3}{*}{$M_2$} & bias & -0.002 & 0.119 & -0.002 & -0.002 & 0.010\\
		& & std & 0.069 & 0.080 & 0.071 & 0.070 & 0.071\\
		& & rmse & 0.069 & 0.143 & 0.071 & 0.070 & 0.072\\
		& \multirow{3}{*}{$M_3$} & bias & -0.000 & 0.045 & -0.001 & -0.001 & 0.008\\
		& & std & 0.039 & 0.044 & 0.042 & 0.043& 0.042\\
		& & rmse & 0.039 & 0.063 & 0.042 & 0.042 & 0.043\\
		\hline
		
		\multirow{6}{*}{$R_3$} & \multirow{3}{*}{$M_1$} & bias & 0.000 & 0.098 & -0.032 & -0.062 & -0.004\\
		& & std & 0.036 & 0.051 & 0.053 & 0.045 & 0.044\\
		& & rmse & 0.036 & 0.110 & 0.062 & 0.076& 0.044 \\
		& \multirow{3}{*}{$M_2$} & bias & -0.001 & 0.095 & -0.016 & -0.036 & -0.004\\
		& & std & 0.068 & 0.090 & 0.071 & 0.069 & 0.071 \\
		& & rmse & 0.068 & 0.130 & 0.073 & 0.078 & 0.071\\
		& \multirow{3}{*}{$M_3$} & bias & -0.001 & 0.065 & -0.001 & -0.010 & 0.006\\
		& & std & 0.038 & 0.053 & 0.045 & 0.047 & 0.045 \\
		& & rmse & 0.038 & 0.084 & 0.045 & 0.048 & 0.045\\
		\hline
	\end{tabular}
\end{table}

\begin{table}
	\centering
		\caption{Simulation results (part II) from $B=2,000$ Monte Carlo studies}\label{tbl2}
	\begin{tabular}{lllrrrrr}
		\hline
		Res & Model & Estimates &  $\theta_{full}$ & $\theta_{CC}$ & $\theta_{KC}$  & $\theta_{FI}$ & $\theta_{SP}$\\
		\hline 
		\multirow{6}{*}{$R_4$} & \multirow{3}{*}{$M_1$} & bias & -0.002 & 0.085 & -0.027 & -0.051 & -0.002\\
		& & std & 0.035& 0.052 & 0.053 & 0.045 & 0.044 \\
		& & rmse & 0.035 & 0.100 & 0.060& 0.068 & 0.044\\
		& \multirow{3}{*}{$M_2$} & bias & 0.001& 0.112 & 0.018 & -0.038& -0.001\\
		& & std & 0.068 & 0.092 & 0.071 & 0.069 & 0.071 \\
		& & rmse & 0.068 & 0.145& 0.073 & 0.079 & 0.071\\
		& \multirow{3}{*}{$M_3$} & bias & -0.002& 0.063 & -0.002 & -0.011& 0.004\\
		& & std & 0.039 & 0.054 & 0.046 & 0.048 & 0.046\\
		& & rmse & 0.039& 0.083 & 0.046 & 0.049 & 0.046\\
		\hline
		\multirow{6}{*}{$R_5$} & \multirow{3}{*}{$M_1$} & bias & -0.000 & 0.092 & -0.029 & -0.055 & -0.002\\
		& & std & 0.036 & 0.051 & 0.054 & 0.045 & 0.044 \\
		& & rmse & 0.036 & 0.105 & 0.061 & 0.071 & 0.045 \\
		& \multirow{3}{*}{$M_2$} & bias & 0.001 & 0.102 & 0.019 & -0.035 & -0.001\\
		& & std & 0.065 & 0.088 & 0.068 & 0.066 & 0.068\\
		& & rmse & 0.065 & 0.134 & 0.071 & 0.074 & 0.068\\
		& \multirow{3}{*}{$M_3$} & bias & -0.001 & 0.063 & -0.001 & -0.010 & 0.007\\
		& & std & 0.038 & 0.053 & 0.046 & 0.048& 0.045\\
		& & rmse & 0.038 & 0.082 & 0.046 & 0.049 & 0.046\\
		\hline
		
		\multirow{6}{*}{$R_6$} & \multirow{3}{*}{$M_1$} & bias & -0.001 & 0.113 & -0.031 & -0.061 & 0.000\\
		& & std & 0.036 & 0.054 & 0.056 & 0.047& 0.045\\
		& & rmse & 0.036 & 0.126 & 0.064 & 0.077& 0.045 \\
		& \multirow{3}{*}{$M_2$} & bias & -0.000 & 0.125 & 0.019 & -0.044 & -0.001\\
		& & std & 0.067 & 0.090 & 0.070 & 0.068 & 0.070 \\
		& & rmse & 0.067 & 0.154 & 0.072 & 0.081 & 0.070\\
		& \multirow{3}{*}{$M_3$} & bias & 0.000 & 0.080 & 0.000 & -0.011 & 0.009\\
		& & std & 0.040 & 0.056 & 0.047 & 0.049 & 0.046 \\
		& & rmse & 0.040 & 0.098 & 0.047 & 0.050 & 0.047\\
		\hline
	\end{tabular}
\end{table}

\begin{table}
	\centering
		\caption{Simulation results (part III) from $B=2,000$ Monte Carlo studies}\label{tbl3}
	\begin{tabular}{lllrrrrr}
		\hline
		Res & Model & Estimates & $\theta_{full}$ & $\theta_{CC}$ & $\theta_{KC}$  & $\theta_{FI}$ & $\theta_{SP}$\\
		\hline 
		\multirow{6}{*}{$R_7$} & \multirow{3}{*}{$M_1$} & bias & -0.000 & 0.092 & 0.020 & -0.038 & -0.001\\
		& & std & 0.068& 0.091 & 0.070 & 0.068 & 0.071 \\
		& & rmse & 0.068 & 0.129 & 0.073& 0.078 & 0.071\\
		& \multirow{3}{*}{$M_2$} & bias & -0.000& 0.092 & 0.020 & -0.038& -0.001\\
		& & std & 0.068 & 0.091 & 0.070 & 0.068 & 0.071 \\
		& & rmse & 0.068 & 0.129& 0.073 & 0.078 & 0.071\\
		& \multirow{3}{*}{$M_3$} & bias & -0.000& 0.071 & -0.001 & -0.011& 0.009\\
		& & std & 0.038 & 0.056 & 0.046 & 0.049 & 0.046\\
		& & rmse & 0.038& 0.090 & 0.046 & 0.050 & 0.047\\
		\hline
		\multirow{6}{*}{$R_8$} & \multirow{3}{*}{$M_1$} & bias & -0.002 & 0.069 & 0.012 & -0.024 & -0.003\\
		& & std & 0.068& 0.086 & 0.070 & 0.068 & 0.070 \\
		& & rmse & 0.068 & 0.110 & 0.071 & 0.072 & 0.070 \\
		& \multirow{3}{*}{$M_2$} & bias & -0.002 & 0.069 & 0.012 & -0.024 & -0.003\\
		& & std & 0.068 & 0.086 & 0.070 & 0.068 & 0.070\\
		& & rmse & 0.068& 0.110 & 0.071 & 0.072 & 0.070\\
		& \multirow{3}{*}{$M_3$} & bias & -0.001 & 0.039 & -0.001 & -0.005 & 0.005\\
		& & std & 0.039 & 0.051 & 0.045 & 0.046& 0.045\\
		& & rmse & 0.039 & 0.064 & 0.045 & 0.046 & 0.045\\
		\hline
		
		\multirow{6}{*}{$R_9$} & \multirow{3}{*}{$M_1$} & bias & 0.002 & 0.099 & 0.016 & -0.036 & -0.001\\
		& & std & 0.069 & 0.089 & 0.071 & 0.069& 0.071\\
		& & rmse & 0.069 & 0.133 & 0.072 & 0.078& 0.071 \\
		& \multirow{3}{*}{$M_2$} & bias & 0.002 & 0.099 & 0.016 & -0.036& -0.001\\
		& & std & 0.069 & 0.089 & 0.071 & 0.069 & 0.071 \\
		& & rmse & 0.069 & 0.133 & 0.072 & 0.078 & 0.071\\
		& \multirow{3}{*}{$M_3$} & bias & 0.000 & 0.066 & -0.009 & -0.018 & 0.002\\
		& & std & 0.039 & 0.055 & 0.046 & 0.047 & 0.046 \\
		& & rmse & 0.039 & 0.086 & 0.046 & 0.051 & 0.046\\
		\hline
	\end{tabular}
\end{table}

From Table \ref{tbl1}, when the response model is logistic linear $(\mathcal{R}_1/ \mathcal{R}_2)$, all methods are consistent. For quadratic model $M_1$, $\theta_{FI}$ and $\theta_{SP}$ are more efficient than $\theta_{KC}$. Under $M_2, M_3$, $\theta_{FI}$ and $\theta_{SP}$ are no worse than $\theta_{KC}$. When the response model is logistic quadratic $(\mathcal{R}_3)$, $\theta_{KC}$ and $\theta_{FI}$ are biased under $M_1$. However, the proposed $\theta_{SP}$ is still consistent and has smaller mean square error. When the outcome regression model is $M_2$, which is slightly violated the linearity, $\theta_{FI}$ is biased and $\theta_{KC}$ is slightly biased. The proposed $\theta_{SP}$ performs better than $\theta_{FI}$ and $\theta_{KC}$ in terms of mean square error. When the outcome regression model is linear $M_3$, $\theta_{SP}$ and $\theta_{KC}$ are consistent, but $\theta_{FI}$ is slightly biased. In terms of efficiency, $\theta_{SP}$ and $\theta_{GMM}$ are better, because $f(Y\mid X, \delta=1)$ uses the full models and induces additional noise from the quadratic terms. 

From table \ref{tbl2}, when the linearity assumption of $X$ in response model is violated, the proposed method still works well. For nonlinear outcome regression models $(M_1/M_2)$, $\theta_{KC}$ and $\theta_{FI}$ are biased due to model misspecification. However, the proposed method is always consistent. For linear outcome regression model $(M_3)$, $\theta_{KC}$ and $\theta_{SP}$ are consistent.

From Table \ref{tbl3}, the misspecification of link function in the response model does not effect the consistency of the proposed method. Furthermore, the violation of the instrumental assumption also does not effect the proposed method heavily. In summary, the proposed method outperforms $\theta_{KC}$ and $\theta_{FI}$. Also, the proposed method suffers less model misspecification.

\subsection{Simulation Study II}
%

In this section, we perform simulation studies to validate the proposed test statistic in Section \ref{sec:test}. The power of the proposed test is related to the non-constant effect of $g(y)$ and sample size. Thus, we design a $4\times 2$ factorial studies, where factors are the coefficient of $g(y)$ and the sample size. 

Assume the superpopulation model is generated as as follows: First, covariate variables $x_i=(x_{i1},x_{i2})$ are generated independently from multivariate normal distribution with mean $(1,1)$ and variance $\text{Diag}(0.25,0.25)$. Second, response variables $y_i$ are generated independently from normal distribution $N(-1+x_{i1}+x_{i2}, 0.25)$. 

Assume the response function is 
\begin{eqnarray}
p_i=\frac{\exp(0.1x_{i1}+\phi_yy_i^2)}{1+\exp(0.1x_{i1}+\phi_yy_i^2)}.\nonumber
\end{eqnarray}
The response indicator functions are generated from a simple random sampling with replacement process with approximate response rate being 70\%. The first order inclusion probabilities are $\{p_i\}_{i=1}^n$.

The whole simulation process can be described as follows:
\begin{itemize}
	\item [1.] Generate the complete sample from the superpopulation model with size $n\in\{100, 500\}$.
	\item [2.] Apply the response mechanism to create nonresponse with $\{0, 0.2, 0.5,1\}$.
	\item[3.] Apply the proposed bootstrap method in Appendix \ref{App:App_B} to obtain the empirical distribution of the proposed test statistic.  
	\item [4.] Repeat step 1--3 $B=1,000$ times.
\end{itemize}


The simulation results are presented in Table \ref{tbl4}.

\begin{table}[ht]
\centering
\caption{Relative number of rejections from $B=1,000$ Monte Carlo studies. $\alpha$ is the predetermined type I error.}\label{tbl4}
\begin{tabular}{l|l|lllll}
\hline
$n$ & $\phi_y$ & \multicolumn{5}{c}{$\alpha$}\\
    &          & 0.01 & 0.05 & 0.1 & 0.15 & 0.2\\
 \hline 
 \multirow{4}{*}{$100$} & 0 & 0 & 0 & 0 & 0 & 0\\
                        & 0.2 & 0.009 & 0.036 & 0.071 & 0.125 & 0.188\\
                        & 0.5 & 0.013 & 0.062 & 0.149 & 0.251 & 0.341\\
                        & 1 & 0.018 & 0.093 & 0.229 & 0.372 & 0.517 \\
 \hline
  \multirow{4}{*}{$500$} & 0 & 0.007 & 0.037 & 0.079 & 0.121 &  0.161\\
 & 0.2 & 0.039 &  0.135 & 0.239 & 0.344 & 0.423\\
 & 0.5 & 0.177 & 0.426 & 0.634 & 0.800 & 0.882\\
 & 1 & 0.344 & 0.705 & 0.888 & 0.980 & 0.995 \\
 \hline
\end{tabular}
\end{table}

The power of the test is that the probability of rejecting the null hypothesis, given that the alternative hypothesis is true. From Table \ref{tbl4}, the power of the proposed test statistic is increasing as the violation $(\phi_y)$ of constant $g(y)$ increases for fixed sample size. For fixed $\phi_y$, the power of the proposed test statistic also increases as sample size increases. For $\phi_y=0$, which indicates the null hypothesis is true, the proposed test statistic can achieves the type I error bound approximately when sample size is $500$. In summary, the proposed test statistic and the bootstrap method can be used to test the ignorability effectively.

\section{Application}\label{sec:application}

\begin{figure}[H]
	\subfigure[The original KLIPS data]{\includegraphics[width=0.6\textwidth]{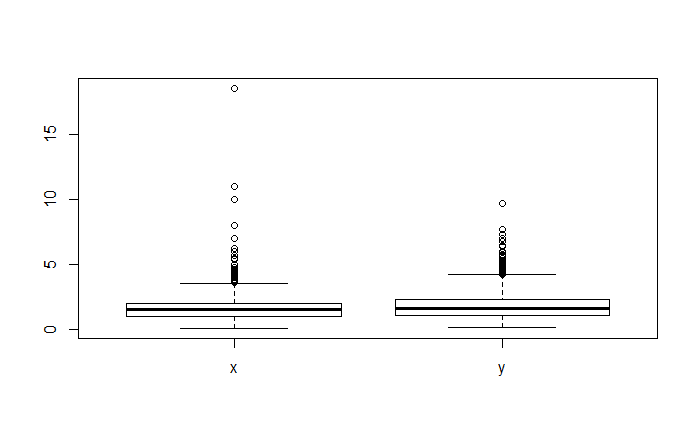}}
	\subfigure[The transformed KLIPS data: $(x,y)\xleftarrow{}\log(x,y)/2$]{\includegraphics[width=0.6\textwidth]{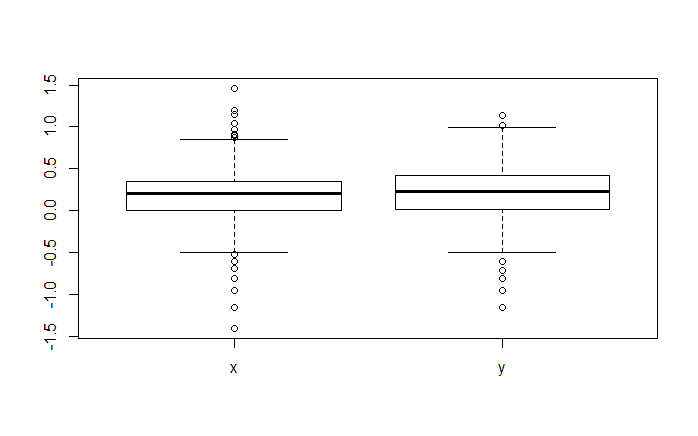}}
	\caption{KLIPS data description ( $\times 10^6$ Korean Won).\label{fig:real_data}}
\end{figure}

In this section, the proposed method is applied to Korea Labor and Income Panel Survey (KLIPS). The introduction of the penal survey can be checked out at \url{http://www.kli.re.kr/klips/en/about/introduce.jsp}. The study variable $(y)$ is the average monthly income for the current year and the auxiliary variable $(x)$ is the average monthly income for the previous year. The KLIPS has $n=2,506$ regular wage earners. And the boxplots for $x$ and $y$ are presented in Figure \ref{fig:real_data}. Note that both $x,y$ has outliers which cause challenging to the nonparametric smoothing method. Thus, we take the transformation to both $x$ and $y$.

Since the KLIPS data are completed, we artificially create the missingness and then apply the proposed method to the incomplete data.  Assume the true response mechanisms are
\begin{eqnarray}
& \mathcal R_1: & Pr(\delta=1\mid x, y)=\left\lbrace 1+\exp(-1+y) \right\rbrace^{-1},\nonumber\\
& \mathcal{R}_2: & Pr(\delta=1\mid x, y)=\left[  1+\exp\left\lbrace -2+\exp(0.5y)\right\rbrace \right] ^{-1}, \nonumber\\
& \mathcal{R}_3: & Pr(\delta=1\mid x, y)=\left\lbrace 
\begin{array}{ll}
0.7 & \text{if $y<0.5$}\\
0.4 & \text{otherwise}
\end{array}\right., \nonumber\\
& \mathcal{R}_4: & Pr(\delta=1\mid x, y)=\Phi\left\lbrace -0.1+0.1\exp(0.5y)\right\rbrace.\nonumber
\end{eqnarray}

The process is described as following:
\begin{itemize}
	\item [1.] Use Simple Random Sampling without Replacement (SRSWOR) to obtain $n$ sample units.
	\item[2.] Apply the response mechanism $\mathcal R$ to the sample and get the incomplete sample.
	\item[3.] Apply the proposed method to the incomplete sample and obtain the parameter estimation.
\end{itemize}

\begin{figure}[H]
	\centering
	\includegraphics[width=1\textwidth]{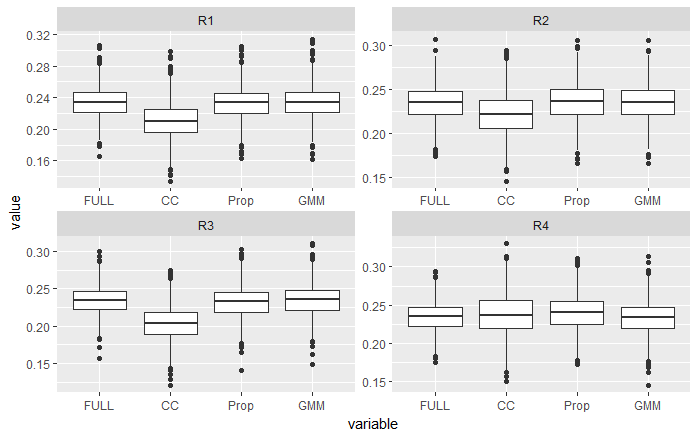}
	\caption{Boxpliots of the estimators for Full, CC, Proposed, and GMM methods. \label{fig:application} }
\end{figure}

Let $n=200$ and replicate the process $B=2,000$ times. For each realized sample, apply Full, CC, Proposed and GMM method to estimate $\theta=E(y)$.
The results are shown in Figure \ref{fig:application}.

From Figure \ref{fig:application}, we can see that both proposed and GMM methods achieve consistent estimates and their efficiencies are comparable. CC methods are always biased. The proposed method is consistent, since it does involve model specifications. The GMM method is consistent in the real data due to the linearity of $x$ and $y$.

\section{Discussion}\label{sec:Disscussion}
In this paper, we propose a profile likelihood method to achieve robust estimation under a semiparametric nonignorable nonresponse model.  From simulation results, our proposed method shows more robustness than generalized linear response models. The proposed method uses the maximum profile likelihood method and an efficient computation algorithm based on fractional imputation is developed. From  asymptotic properties, our proposed method enjoys $\sqrt{n}$-consistency. Furthermore, our proposed method assumes the response mechanism is a flexible function of $Y$.  Then, we propose a test procedure to check if the response mechanism is missing at random. The bootstrap method is proposed to obtain the empirical distribution of the proposed test statistic. Our proposed method can be used in survey data directly by replacing the likelihood function to the pseudo likelihood function.

\appendix
\clearpage
\numberwithin{equation}{section}

\section{Derivations in M-Step}\label{App:App_A}
	Note that, $\tilde{l}_{obs}(\phi, g\mid w^{*(t)})$ are generalized partially linear function of $\phi$ and $g$. Then, the profile method likelihood can be applied. The outlined procedures are described as follows.
First, $g(y)$ can be estimated by maximizing
\begin{eqnarray}
&\tilde{l}_{obs}(\phi, g\mid w^{*(t)})&=\sum_{i=1}^{n} \delta_i\log \pi\left\lbrace x_{i1}^T\phi+g(y)\right\rbrace K_h(y-y_i)\nonumber\\
&&+(1-\delta_i)\sum_{j=1}^M w_{ij}^{*(t)}\log \left[ 1-\pi\left\lbrace  x_{i1}^T\phi+g(y)\right\rbrace  \right]  K_h(y-y_{ij}^*),\nonumber
\end{eqnarray}
given a fixed $\phi$. Denote it as $\hat g_{\phi}(y)$.
Then, $\phi$ can be estimated by maximizing 
\begin{eqnarray}
\hat l_{obs}(\phi, \hat g_{\phi}\mid w^{*(t)})=\sum_{i=1}^{n} \delta_i \log \pi\left\lbrace  x_{i1}^T\phi+\hat g_{\phi}(y_i)\right\rbrace + (1-\delta_i)\sum_{j=1}^M w_{ij}^{*(t)}\log \left[1- \pi\left\lbrace  x_{i1}^T\phi+\hat g_{\phi}(y_{ij}^*) \right\rbrace\right] .\nonumber
\end{eqnarray}

The details of one-step Newton-Raphson algorithm are shown as follows.
 The maximization of $\tilde{l}_{obs}(\phi, g\mid w^{*(t)})$ respect to $g(y)$ is equivalent to taking the first order derivative respect to $g(y)$. That is
\begin{eqnarray}
&\frac{\partial \tilde{l}_{obs}(\phi, g\mid w^{*(t)})}{\partial g(y)}&=\sum_{i=1}^{n}\delta_i \left[ 1-\pi\left\lbrace  x_{i1}^T\phi+g(y)\right\rbrace\right] K_h(y-y_i)\nonumber\\
&&-(1-\delta_i)\sum_{j=1}^M w_{ij}^{*(t)}\pi\left\lbrace x_{i1}^T\phi+g(y)\right\rbrace K_h(y-y_{ij}^*).\nonumber\label{gradient_of_lh}
\end{eqnarray}
To estimate $g(y)$, it is equivalent to solving $\partial \tilde{l}_{obs}(\phi, g\mid w^{*(t)})/\partial g(y)=0$. Applying the one-step Newton-Raphson, we can update the estimator by
\begin{eqnarray}
g(y)^{(t+1)}=g^{(t)}(y)-\frac{G_t(y)}{H_t(y)}\nonumber
\end{eqnarray}
where
$$ G_t(y)=\sum_{i=1}^{n}\delta_i \left[ 1-\pi\left\lbrace x_{i1}^T\phi^{(t)}+g^{(t)}(y)\right\rbrace\right] K_h(y-y_i)-(1-\delta_i)\sum_{j=1}^M w_{ij}^{*(t)}\pi\left\lbrace  x_{i1}^T\phi^{(t)}+g^{(t)}(y)\right\rbrace K_h(y-y_{ij}^*)$$ is the gradient of $\tilde{l}_{obs}(\phi, g\mid w^{*(t)})$ respect to $g(y)$, and 
\begin{eqnarray}
&H_t(y)&= -\sum_{i=1}^{n}\left[ 1-\pi\left\lbrace x_{i1}^T\phi^{(t)}+g^{(t)}(y)\right\rbrace\right]\pi\left\lbrace  x_{i1}^T\phi^{(t)}+g^{(t)}(y)\right\rbrace\nonumber\\
&&\times \left\lbrace \delta_iK_h(y-y_i) +(1-\delta_i)\sum_j^M w_{ij}^{*(t)}K_h(y-y_{ij}^*)\right\rbrace,\nonumber
\end{eqnarray}
is the Hessian matrix of $\tilde{l}_{obs}(\phi, g\mid w^{*(t)})$ respect to $g(y)$.

Note that $g(y)$ is the function of $\phi$. Thus, take the partial derivative of $\tilde{l}_{obs}(\phi, g\mid w^{*(t)})/\partial g(y)$ respect to $\phi$ and set it to be 0. That is 
\begin{eqnarray}
&\frac{\partial^2\tilde{l}_{obs}(\phi, g\mid w^{*(t)})}{\partial g(y)\partial \phi}&=-\sum_{i=1}^{n}\left[ 1-\pi\left\lbrace x_{i1}^T\phi+g(y)\right\rbrace\right]\pi\left\lbrace x_{i1}^T\phi+g(y)\right\rbrace\nonumber\\
&&\left\lbrace \delta_iK_h(y-y_i) +(1-\delta_i)\sum_j^M w_{ij}^{*(t)}K_h(y-y_{ij}^*)\right\rbrace \left\lbrace x_{i1}+\bigtriangledown g(y) \right\rbrace=0,\nonumber
\end{eqnarray}
where $\bigtriangledown g(y)=\frac{\partial g(y)}{\partial \phi}$.  Solving $\partial^2\tilde{l}_{obs}(\phi, g\mid w^{*(t)})/\left\lbrace \partial g(y)\partial \phi\right\rbrace =0$, we can obtain a closed form for $\bigtriangledown g(y)$ as
\begin{eqnarray}
\bigtriangledown g^{(t)}(y)=\frac{I_t(y)}{H_t(y)},\label{gradient_of_g}\nonumber
\end{eqnarray}
where 
\begin{eqnarray}
 &I_t(y)&=\sum_{i=1}^{n}\left[ 1-\pi\left\lbrace x_{i1}^T\phi^{(t)}+g^{(t)}(y)\right\rbrace\right]\pi\left\lbrace x_{i1}^T\phi^{(t)}+g^{(t)}(y)\right\rbrace\nonumber\\
 &&\times\left\lbrace \delta_iK_h(y-y_i) +(1-\delta_i)\sum_j^M w_{ij}^{*(t)}K_h(y-y_{ij}^*)\right\rbrace x_{i1}.\nonumber
\end{eqnarray}
Then, $\phi$ can be estimated by maximizing 
\begin{eqnarray}
\hat l_{obs}(\phi, g_{\phi}\mid w^{*(t)})=\sum_{i=1}^{n} \delta_i \log \pi\left\lbrace x_{i1}^T\phi+g_{\phi}(y_i)\right\rbrace + (1-\delta_i)\sum_{j=1}^M w_{ij}^{*(t)}\log \left[1- \pi\left\lbrace x_{i1}^T\phi+g_{\phi}(y_{ij}^*) \right\rbrace\right] ,\nonumber
\end{eqnarray}
which leads to solving
\begin{eqnarray}
\frac{\hat l_{obs}(\phi, g_{\phi}\mid w^{*(t)})}{\partial \phi}=0.\label{A4}\nonumber
\end{eqnarray}
Let 
\begin{eqnarray}
&A_t=\bigtriangledown \hat l_{obs}(\phi, g_{\phi}\mid w^{*(t)})&=\sum_{i=1}^{n}\delta_i\left[ 1-\pi\left\lbrace x_{i1}^T\phi^{(t)}+g^{(t)}(y_i)\right\rbrace\right]\left( x_{i1}+\bigtriangledown g^{(t)}(y_i) \right)\nonumber\\
&&-(1-\delta_i)\sum_{j=1}^{M} w_{ij}^{*(t)}\pi\left\lbrace x_{i1}^{T}\phi^{(t)}+g^{(t)}(y_{ij}^*)\right\rbrace\left( x_{i1}+\bigtriangledown g^{(t)}(y_{ij}^*) \right).\nonumber
\end{eqnarray}
To compute the Hessian matrix of $\hat l_{obs}(\phi, g_{\phi}\mid w^{*(t)})$, we consider $\bigtriangledown g$ to be constant with respect to $\phi$ \citep{muller2001estimation}. This leads to
\begin{eqnarray}
&B_t=&\bigtriangleup \hat l_{obs}(\phi, g_{\phi}\mid w^{*(t)})=-\sum_{i=1}^{n} \delta_i\pi\left\lbrace x_{i1}^T\phi^{(t)}+g^{(t)}(y_i)\right\rbrace \left[ 1-\pi\left\lbrace  x_{i1}^T\phi^{(t)}+g^{(t)}(y_i)\right\rbrace\right]\nonumber\\
&&\times\left( x_{i1}+\bigtriangledown g^{(t)}(y_i) \right)^{\otimes 2}
+(1-\delta_i)\sum_{j=1}^{M}w_{ij}^{*(t)}\pi\left\lbrace x_{i1}^T\phi^{(t)}+g^{(t)}(y_{ij}^*)\right\rbrace\left[ 1-\pi\left\lbrace x_{i1}^T\phi^{(t)}+g^{(t)}(y_{ij}^*)\right\rbrace\right]\nonumber\\
&&\times\left( x_{i1}+\bigtriangledown g^{(t)}(y_{ij}^*) \right)^{\otimes 2},\nonumber
\end{eqnarray}
where $A^{\otimes 2}=AA^T$. Thus, applying Newton-Raphson algorithm, we can update $\phi$ by 
\begin{eqnarray}
\phi^{(t+1)}=\phi^t-B_t^{-1}A_t.\nonumber
\end{eqnarray}

\section{Algorithm for Bootstrap}\label{App:App_B}
From the proposed method in \S \ref{sec:proposal}, a pseudo complete sample $\{(x_i, \hat y_i, \delta_i)\}_{i=1}^n$ can be obtained, where
\begin{eqnarray}
\hat y_i=\left\lbrace 
\begin{array}{ll}
 y_i & \text{if $\delta_i=1$}\\
 \sum_{j=1}^{M}w_{ij}^*y_{ij}^* & \text{otherwise}.
\end{array}\right.\nonumber
\end{eqnarray}

As discussed in \S \ref{sec:test}, under the null hypothesis, $(\hat \phi_a, \hat c_a)$ can be obtained by maximizing (\ref{like_mar}). Then, the proposed parametric bootstrap can be described as follows:
\begin{itemize}
	\item []\textit{Step 1}: Using $(\hat \phi_a, \hat c_a)$, we can regenerate the response indicators $\delta_i^*$ from the Bernoulli distribution with success probability $\pi(\hat \phi_a, \hat c_a; x_i)$. Then, we can formulate the new pseudo sample $\{x_i, \delta_i^* \hat y_i, \delta_i^*\}_{i=1}^n$. 
	\item[] \textit{Step 2}: Apply $\{x_i, \delta_i^* \hat y_i, \delta_i^*\}_{i=1}^n$ to (\ref{like_mar}) to obtain $(\hat \phi_a^*, \hat c_a^*)$. 
	\item[] \textit{Step 3}: Apply $\{x_i, \delta_i^* \hat y_i, \delta_i^*\}_{i=1}^n$ to the proposed method and compute the test statistic $\hat R^k$ in (\ref{test2}).
	\item[] \textit{Step 4}: Repeat \textit{Step 1--3} $B$ times and compute the p-value as 
	\begin{eqnarray}
	\text{p-value}=\frac{1}{B}\sum_{k=1}^{B}I(\hat B<\hat B^k).\nonumber
	\end{eqnarray}
\end{itemize}

If the p-value is less than the type I error $\alpha$, then we reject $\mathrm{H}_0$. Otherwise, we have no significant evidence to reject $\mathrm{H}_0$.

\section{Regularity conditions and Proof of Lemma \ref{Monotone} and Theorem \ref{thm:thm1}}\label{App:App_C}

Regularity conditions of (\textbf{C3}) are described as follows.
\begin{itemize}
	\item []\textit{C3(a)}: For $\eta$ in an open subset, assume $s(\eta; X, Y)$ is twice continuously differentiable for every $X,Y$.
	\item[]\textit{C3(b)}: Assume there exists $\eta_0$, such that $E\left\lbrace s(\eta_0; X, Y) \right\rbrace=0 $.
	\item[] \textit{C3(c)}: For $\eta$ in a neighborhood of $\eta_0$, assume $E\left\lbrace \|s(\eta; X, Y)\|^2\right\rbrace<\infty $ and $E\left\lbrace \partial s(\eta; X, Y)/\partial \eta^T\right\rbrace $ exists and is nonsingular.
\end{itemize}

Regularity conditions of (\textbf{C5}) are described as follows.
\begin{itemize}
	\item []\textit{C5(a)}: The response probability $\pi(X, Y)$ is bonded below from 0 uniformly.
	\item[]\textit{C5(b)}: There exists $\theta_0$, such that $E\left\lbrace U(\theta_0; X,Y) \right\rbrace=0 $.
	\item[]\textit{C5(c)}: For $\theta$ in a neighborhood of $\theta_0$, assume $U(\theta; X,Y)$ is twice continuously differentiable for every $X, Y$.
	\item[] \textit{C5(d)}: For $\theta$ in a neighborhood of $\theta_0$, assume $E\left\lbrace \|U(\theta; X< Y)\|^2\right\rbrace <\infty$ and $E\left\lbrace \partial U(\theta; X, Y)/\partial \theta^T\right\rbrace $ exists and is nonsingular.
\end{itemize}

The road map of this proof can be outlined as follows.
\begin{itemize}
	\item []\textit{Step 1}: We will show the asymptotic normality of the profile estimator of $\beta$ under complete data using 
	\begin{eqnarray}
	l_{Full}(\phi, g)=\sum_{i=1}^{n}\left( \delta_i\log  \pi\left\lbrace \phi; x_{i1},g(y_i)\right\rbrace +(1-\delta_i)\log \left[  1- \pi\left\lbrace \phi;  x_{i1},g(y_i)\right\rbrace \right]\right)  .\nonumber
	\end{eqnarray} 
	\item[] \textit{Step 2}: Then, we can establish the asymptotic distribution under nonresponse using
	\begin{eqnarray}
	&l_{obs}(\phi, g;\eta_0)=&\sum_{i=1}^{n}\left[ \delta_i\log  \pi\left\lbrace \phi; x_{i1},g(y_i)\right\rbrace\right.\nonumber\\
	&&\left. +(1-\delta_i)E\left(  \log \left[  1- \pi\left\lbrace \phi; x_{i1},g(y)\right\rbrace\right] \mid x_i, \delta_i=0; \eta_0\right)\right] \nonumber.
	\end{eqnarray}
	\item[] \textit{Step 3}: The asymptotic distribution is further extended to incorporate the estimation of $\eta_0$.
	\item[] \textit{Step 4}: Finally, we will show that the proposed algorithm is equivalent to  applying the profile method to $l_{obs}(\phi, g;\hat \eta)$ asymptotically. 
\end{itemize}

Let us first show \textit{Step 1}.  Since $g$ maps a scalar $y$ into some space $G$, define $\zeta=g(y)\in G$. Let
\begin{eqnarray}
p(\delta; \phi, \zeta)=\pi\left\lbrace\phi, \zeta; x_1,y \right\rbrace^{\delta} \left[ 1-\pi\left\lbrace\phi, \zeta; x_1,y \right\rbrace\right]^{1-\delta} \nonumber
\end{eqnarray}
as the conditional distribution of $\delta$ given $(x, y)$. Furthermore, let $l(\delta; \phi, \zeta)=\log p(\delta; \phi, \zeta) $.
Let $\hat g_{\phi}$ be the solution of maximizing
\begin{eqnarray}
	\tilde l_{Full}(\phi, g)=\sum_{i=1}^{n}\left(  \delta_i\log  \pi\left\lbrace \phi; x_{i1},g(y)\right\rbrace +(1-\delta_i)\log \left[  1- \pi\left\lbrace \phi;  x_{i1},g(y)\right\rbrace \right]\right) K_h(y-y_i).\nonumber
\end{eqnarray}
Let $\hat \phi$ be the maximizer of $l_{Full}(\phi, \hat g_{\phi})$.
Furthermore, we define the Fr\'echet derivative of $l_{Full}(\phi, g)$ respect to function $g$ as
\begin{eqnarray}
\frac{\partial l_{Full}(\phi, g)}{\partial g}=\left.\frac{\partial l_{Full}(\phi, g+\lambda u)}{\partial \lambda}\right|_{\lambda=0}.\nonumber
\end{eqnarray}

Following the proof in \cite{severini1992profile}, we present the sufficient conditions to obtain the asymptotic distribution. 
\begin{itemize}
	\item[]\textit{Assumption 1.} For any fixed $\phi_1 \in \Phi$ and $\zeta_1 \in G$, let 
	\begin{eqnarray}
	\rho(\phi, \zeta)=\int \log p(\delta;\phi, \zeta) p(\delta; \phi_1, \zeta_1) \mathrm{d}\delta.\nonumber
	\end{eqnarray}
	If $\phi\neq\phi_1$, then
	\begin{eqnarray}
	\rho(\phi, \zeta)<\rho(\phi_1, \zeta_1).\nonumber
	\end{eqnarray}
	\item[] \textit{Assumption 2.} Define the marginal Fisher information for $\phi$ as
	\begin{eqnarray}
	\tilde{I}_{\phi}(\phi, \zeta)=E_{\phi, \zeta}\left\lbrace \frac{\partial l}{\partial \phi}(\delta; \phi, \zeta)^2 \right\rbrace -E_{\phi, \zeta}\left\lbrace \frac{\partial l}{\partial \phi}(\delta; \phi, \zeta)\frac{\partial l}{\partial \zeta}(\delta; \phi, \zeta)\right\rbrace^2  E_{\phi, \zeta}\left\lbrace \frac{\partial l}{\partial \zeta}(\delta; \phi, \zeta)^2 \right\rbrace^{-1}.\nonumber
	\end{eqnarray}
	Assume $\tilde{I}_{\phi}(\phi, \zeta)>0$ for all $\phi\in \Phi$ and $\zeta \in G$.
	
	\item[] \textit{Assumption 3.} Assume that the derivative
	\begin{eqnarray}
	\frac{\partial^{r+s}l}{\partial \phi^r\partial \zeta^s}l(\delta; \phi, \zeta)\nonumber
	\end{eqnarray}
	exists for all $r\geq 0,s\geq 0, r+s\leq 4$.  Moreover, 
	\begin{eqnarray}
	E_0\left\lbrace \sup_{\phi}\sup_{\zeta} \left\| \frac{\partial^{r+s}l}{\partial \phi^r\partial \zeta^s}l(\delta; \phi, \zeta)\right\|^2 \right\rbrace \leq \infty,\nonumber
	\end{eqnarray}
	where $E_0$ denotes expectation under the true density function.
	\item[]\textit{Assumption 4.} Assume the unction $g(y)$ satisfies the Conditions NP (Nuisance parameter) in \cite{severini1992profile}.
\end{itemize}

The following lemma is established from \cite{severini1992profile} and we are using the special case of logistic semiparametric model. 
\begin{lemma}\label{lemmaC1}
	Under Assumption 1--4, we can show\begin{eqnarray}
	\sqrt{n}(\hat \phi-\phi_0)\xrightarrow{}N(0, \tilde{I}_{\phi_0}^{-1}),\nonumber
	\end{eqnarray}
	where $\tilde{I}_{\phi_0}$ is the marginal Fisher information for $\phi_0$. Then, we can also establish that 
	\begin{eqnarray}
	\frac{1}{\sqrt n} \left.\frac{d}{d\phi} \frac{\partial 	l_{Full}(\phi, g_{\phi}) }{\partial g}\right|_{\phi=\phi_0} (\hat g_0-g_0)=o_p(1),\nonumber\\
	\frac{1}{\sqrt n} \left. \frac{\partial 	l_{Full}(\phi, g_{\phi}) }{\partial g}\right|_{\phi=\phi_0} (\hat g'_0-g'_0)=o_p(1),\nonumber
	\end{eqnarray}
	where $g_0=g_{\phi_0}$ is the true function, $\hat g_0=\hat g_{\phi_0}$ and $g'=\frac{dg(y)}{dy}$. 
\end{lemma}
This completes \textit{Step 1}. \textit{Step 1} is a standard conclusion from \cite{severini1992profile}.

Then, we want to extent Lemma (\ref{lemmaC1}) to nonresponse. Note that $l_{obs}(\phi, g;\eta_0)=E\left\lbrace l_{Full}(\phi, g)\mid X, Y_{obs}, R; \eta_0\right\rbrace $, where $X=(x_1, x_2, \cdots, x_n)$, $Y_{obs}$ is the observed part of $(y_1, \cdots, y_n)$ and $R=(\delta_1, \cdots, \delta_n)$. Similarly, the smoothed observed log-likelihood is $\tilde{l}_{obs}(\phi, g; \eta_0)=E\left\lbrace \tilde l_{Full}(\phi, g)\mid X, Y_{obs}, R; \eta_0\right\rbrace$. Then, we can establish the following lemma.
\begin{lemma}\label{lemmaC2}
	Let $\hat g_{\phi}$ be the maximizer of $\tilde l_{Full}(\phi, g)$, then
 $	\hat g_{\phi, obs}=E(\hat g_{\phi}\mid X, Y_{obs}, R; \eta_0) $ is the maximizer of $\tilde{l}_{obs}(\phi, g; \eta_0)$.
\end{lemma}
The proof can be briefly shown as follows. We can use the Fr\'echet derivative and expanse
\begin{eqnarray}
&\tilde l_{Full}(\phi, g)\cong&\tilde l_{Full}(\phi, \hat g_{\phi})+\left.\frac{\partial \tilde l_{Full}(\phi, g)}{\partial g}\right|_{g=\hat g_{\phi}}(g-\hat g_{\phi})+\left.\frac{\partial^2 \tilde l_{Full}(\phi, g)}{\partial g^2}\right|_{g=\hat g_{\phi}}(g-\hat g_{\phi})^2\nonumber\\
&&=\tilde l_{Full}(\phi, \hat g_{\phi})+\left.\frac{\partial^2 \tilde l_{Full}(\phi, g)}{\partial g^2}\right|_{g=\hat g_{\phi}}(g-\hat g_{\phi})^2.\nonumber
\end{eqnarray}
Taking the conditional expectation to both sides, we can obtain that
\begin{eqnarray}
&\tilde{l}_{obs}(\phi, g; \eta_0)\cong &E\left\lbrace \tilde l_{Full}(\phi, \hat g_{\phi})\mid X, Y_{obs}, R; \eta_0\right\rbrace\nonumber\\
&&+E\left\lbrace \left.\frac{\partial^2 \tilde l_{Full}(\phi, g)}{\partial g^2}\right|_{g=\hat g_{\phi}} \mid X, Y_{obs}, R; \eta_0\right\rbrace E\left\lbrace (g-\hat g_{\phi})^2\mid X, Y_{obs}, R; \eta_0\right\rbrace.\nonumber 
\end{eqnarray}
The above equation is upper-bounded at $\hat g_{\phi, obs}$. Then, we complete the proof of Lemma \ref{lemmaC2}.

Then, denote $\hat \phi_{obs}$ be the solution of maximizing $$\tilde{l}_{obs}(\phi,\hat g_{\phi}; \eta_0 )=E\left\lbrace \tilde{l}_{full}(\phi,\hat g_{\phi}; \eta_0 )\mid X, Y_{obs}, R; \eta_0 \right\rbrace .$$

Using Lemma \ref{lemmaC1} and following the same procedures in \cite{severini1992profile}, we can show that \ref{lemmaC1} also holds for $\tilde{l}_{obs}(\phi, g; \zeta_0)$, in the sense of 
\begin{lemma}\label{lemmaC3}
	Assume $\inf_{\phi, g, x, y}\pi(\phi, g; x_1, y)>0$. Under the same assumptions in Lemma \ref{lemmaC1}, we can show that 
	show
	\begin{eqnarray}
	\sqrt{n}(\hat \phi_{obs}-\phi_0)\xrightarrow{}N(0, \tilde{I}_{obs}^{-1}),\nonumber
	\end{eqnarray}
	where $\tilde{I}_{obs}$ is the marginal Fisher information for $\phi_0$ using the observed log-likelihood function. Then, we can also establish that 
	\begin{eqnarray}
	\frac{1}{\sqrt n} \left.\frac{d}{d\phi} \frac{\partial 	l_{obs}(\phi, g_{\phi}) }{\partial g}\right|_{\phi=\phi_0} (\hat g_0-g_0)=o_p(1).\nonumber
	\end{eqnarray}
\end{lemma}
This completes \textit{Step 2}.

Note that $\hat \phi_{obs}$ in Lemma (\ref{lemmaC2}) is a function of $\eta_0$ and we can denote it as $\hat \phi_{obs}(\eta_0)$. However, our profiled estimation is applied to $\tilde{l}_{obs}(\phi,\hat g_{\phi, obs}; \hat \eta )$, where $\hat \eta$ is a solution of
\begin{eqnarray}
U(\eta)=\sum_{i=1}^n\delta_is(\eta; x_i,y_i)=0.\nonumber
\end{eqnarray}
Under the regularity conditions of Z-statistics in \cite{van1998asymptotic}, we can establish that 
\begin{eqnarray}
\sqrt{r}(\hat \eta-\eta_0)\xrightarrow{}N(0, S),\label{eta}
\end{eqnarray}
in distribution,
where $r=\sum_{i=1}^n \delta_i$ and 
\begin{eqnarray}
r \left\lbrace \frac{\partial U(\eta)}{\partial \eta^T}\right\rbrace^{-1}\mathrm{var}\left\lbrace U(\eta)\right\rbrace  \left[ \left\lbrace \frac{\partial U(\eta)}{\partial \eta^T}\right\rbrace^{-1}\right]^T\xrightarrow{}S\nonumber 
\end{eqnarray}
in probability.

To obtain the limiting distribution of $\hat \phi_{obs}(\hat \eta)$, militarization can be used. 
\begin{eqnarray}
\hat \phi_{obs}(\hat \eta)\cong \hat \phi_{obs}(\eta_0)+\frac{\hat \phi_{obs}(\eta_0)}{\partial \eta_0} (\hat\eta- \eta_0).\nonumber
\end{eqnarray}
Moreover, $\hat \phi_{obs}(\eta_0)$ is the solution of
\begin{eqnarray}
\frac{\partial l_{obs}(\phi, \hat g_{\phi,obs}; \eta_0)}{\partial \phi}=0.\nonumber
\end{eqnarray}
Using the derivative of implicit function, we can obtain that
\begin{eqnarray}
\frac{\partial \hat \phi(\eta_0)}{\partial \eta_0}=\left.-\left\lbrace \frac{\partial^2 l_{obs}(\phi, \hat g_{\phi,obs}; \eta_0)}{\partial\phi \partial \phi^T} \right\rbrace^{-1} \frac{\partial^2 l_{obs}(\phi, \hat g_{\phi,obs}; \eta_0)}{\partial \phi\partial \eta_0^T}\right|_{\phi=\hat \phi_{obs}(\eta_0)}.\nonumber
\end{eqnarray}
Furthermore, 
\begin{eqnarray}
\left.-\left\lbrace \frac{\partial^2 l_{obs}(\phi, \hat g_{\phi,obs}; \eta_0)}{\partial\phi \partial \phi^T} \right\rbrace^{-1}\right|_{\phi=\hat \phi_{obs}(\eta_0)} \xrightarrow{}n^{-1}\tilde{I}_{obs}^{-1}\nonumber
\end{eqnarray}
in probability. 
Let 
\begin{eqnarray}
\hat C_n=\left.\frac{\partial^2 l_{obs}(\phi, \hat g_{\phi,obs}; \eta_0)}{\partial \phi\partial \eta_0^T}\right|_{\phi=\hat \phi_{obs}(\eta_0)}=O_p(n).\nonumber
\end{eqnarray}
Thus, we have 
\begin{eqnarray}
\hat \phi_{obs}(\hat \eta)\cong \hat \phi_{obs}(\eta_0)+n^{-1}\tilde{I}_{obs}^{-1} \hat C_n (\hat\eta- \eta_0).\label{phi_eta}
\end{eqnarray}

Combining (\ref{eta}) and (\ref{phi_eta}), we have 
\begin{eqnarray}
\hat \phi_{obs}(\hat \eta)\xrightarrow{}\phi_0,\label{phi0}
\end{eqnarray}
in probability, since $\hat \phi_{obs}(\eta_0)\xrightarrow{}\phi_0$, $n^{-1}\tilde{I}_{obs}^{-1} \hat C_n=O_p(1)$ and $\hat \eta-\eta_0=o_p(1)$.
Then, we can decompose the variance of $\hat \phi_{obs}(\hat \eta)$ as
\begin{eqnarray}
& n\mathrm{var}\left\lbrace \hat \phi_{obs}(\hat \eta)\right\rbrace &\cong n\mathrm{var}\left\lbrace \hat \phi_{obs}(\eta_0)+n^{-1}\tilde{I}_{obs}^{-1} \hat C_n (\hat\eta- \eta_0)\right\rbrace\nonumber\\
&&\cong \tilde{I}_{obs}^{-1}+n^{-1}r^{-1}\tilde{I}_{obs}^{-1} \hat C_n S \hat C_n^T\tilde{I}_{obs}^{-1}+2n\mathrm{Cov}\left\lbrace \hat \phi_{obs}(\eta_0), n^{-1}\tilde{I}_{obs}^{-1} \hat C_n (\hat\eta- \eta_0)\right\rbrace \nonumber\\
&&\xrightarrow{}\tilde{I}_{obs}^{-1}+\Sigma_2+\Sigma_3,\label{var}
\end{eqnarray}
in probability.

Using (\ref{phi0}) and (\ref{var}), we can show that 
\begin{eqnarray}
\sqrt{n}\left\lbrace \hat \phi_{obs}(\hat\eta)-\eta_0\right\rbrace\xrightarrow{}N(0, \Sigma_1), \label{hat_phi}
\end{eqnarray}
where $\Sigma_1=\tilde{I}_{obs}^{-1}+\Sigma_2+\Sigma_3$. This completes \textit{Step 3}.

Define
	\begin{eqnarray}
\hat l_{obs}(\phi, g\mid w^{*(t)})=\sum_{i=1}^{n}\left(  \delta_i \log \pi\left\lbrace \phi; x_{i1},g(y_i)\right\rbrace + (1-\delta_i)\sum_{j=1}^M w_{ij}^{*(t)}\log \left[1- \pi\left\lbrace \phi; x_{i1},g(y_{ij}^*)\right\rbrace  \right]\right).
\end{eqnarray}
The smoothed function is 
\begin{eqnarray}
&\tilde l_{obs}(\phi, g\mid w^{*(t)})=\sum_{i=1}^{n}&\left( \delta_i \log \pi\left\lbrace \phi; x_{i1},g(y)\right\rbrace K_h(y-y_i) \right.\nonumber\\
&&\left.+ (1-\delta_i)\sum_{j=1}^M w_{ij}^{*(t)}\log \left[1- \pi\left\lbrace \phi; x_{i1},g(y_{ij}^*)\right\rbrace  \right]K_h(y-y_{ij}^*)\right).\nonumber
\end{eqnarray}
In our proposed algorithm, \textit{M-Step} is to implement one-step Newton-Raphson method. Finally, we show the following lemma.
 \begin{lemma}
 	For our proposed algorithm, we have 
 	\begin{eqnarray}
 	\hat l_{obs}(\phi^{(t)}, g_{\phi^{(t)}}\mid w^{*(t)})\leq \hat l_{obs}(\phi^{(t+1)}, g_{\phi^{(t+1)}}\mid w^{*(t)})\nonumber,\\
 \tilde l_{obs}(\phi^{(t+1)}, g^{(t)}\mid w^{*(t)})\leq\tilde l_{obs}(\phi^{(t+1)}, g^{(t+1)}\mid w^{*(t)}).\nonumber
 	\end{eqnarray}
 \end{lemma}
Given $w^{*(t)}$, the implementation of \textit{M-step} is 
\begin{eqnarray}
\phi^{(t+1)}=\phi^{(t)}-\left.\left\lbrace \frac{\partial^2 \hat l_{obs}(\phi, g_{\phi}\mid w^{*(t)})}{\partial \phi\partial \phi^T }\right\rbrace^{-1}  \frac{\partial\hat l_{obs}(\phi, g_{\phi}\mid w^{*(t)})}{\partial \phi}\right|_{\phi=\phi^{(t)}},\label{M1}\\
g^{(t+1)}=g^{(t)}-\left.\left\lbrace \frac{\partial^2\tilde  l_{obs}(\phi^{(t+1)}, g\mid w^{*(t)})}{\partial g^2}\right\rbrace^{-1}  \frac{\partial\tilde l_{obs}(\phi^{(t+1)}, g\mid w^{*(t)})}{\partial g}\right|_{g=g^{(t)}}.\label{M2}
\end{eqnarray}
Note that, 
\begin{eqnarray}
&\hat l_{obs}(\phi^{(t+1)}, g_{\phi^{(t+1)}}\mid w^{*(t)})=&\hat l_{obs}(\phi^{(t)}, g_{\phi^{(t)}}\mid w^{*(t)})+\frac{\partial \hat l_{obs}(\phi^{(t)}, g_{\phi^{(t)}}\mid w^{*(t)})}{\partial \left( \phi^{(t)}\right)^T }(\phi^{(t+1)}-\phi^{(t)})\nonumber\\
&&+\frac{1}{2}(\phi^{(t+1)}-\phi^{(t)})^T \frac{\partial^2 \hat l_{obs}(\phi^{(t)}, g_{\phi^{(t)}}\mid w^{*(t)})}{\partial \phi^{(t)} \partial \left( \phi^{(t)}\right)^T }(\phi^{(t+1)}-\phi^{(t)})\nonumber\\
&&+o_p(\|\phi^{(t+1)}-\phi^{(t)}\|^2).\label{t1}
\end{eqnarray}
Plugging (\ref{M1}) into (\ref{t1}), we can obtain 
\begin{eqnarray}
&\hat l_{obs}(\phi^{(t+1)}, g_{\phi^{(t+1)}}\mid w^{*(t)})=&\hat l_{obs}(\phi^{(t)}, g_{\phi^{(t)}}\mid w^{*(t)})\nonumber\\
&&-\frac{1}{2}\left.\frac{\partial\hat l_{obs}(\phi, g_{\phi}\mid w^{*(t)})}{\partial \phi^T}\left\lbrace \frac{\partial^2 \hat l_{obs}(\phi, g_{\phi}\mid w^{*(t)})}{\partial \phi\partial \phi^T }\right\rbrace^{-1}  \frac{\partial\hat l_{obs}(\phi, g_{\phi}\mid w^{*(t)})}{\partial \phi}\right|_{\phi=\phi^{(t)}}.\nonumber
\end{eqnarray}

Since 
\begin{eqnarray}
\left.\frac{\partial\hat l_{obs}(\phi, g_{\phi}\mid w^{*(t)})}{\partial \phi^T}\left\lbrace \frac{\partial^2 \hat l_{obs}(\phi, g_{\phi}\mid w^{*(t)})}{\partial \phi\partial \phi^T }\right\rbrace^{-1}  \frac{\partial\hat l_{obs}(\phi, g_{\phi}\mid w^{*(t)})}{\partial \phi}\right|_{\phi=\phi^{(t)}}\leq 0,\nonumber
\end{eqnarray}
we have 
\begin{eqnarray}
\hat l_{obs}(\phi^{(t)}, g_{\phi^{(t)}}\mid w^{*(t)})\leq \hat l_{obs}(\phi^{(t+1)}, g_{\phi^{(t+1)}}\mid w^{*(t)})
\end{eqnarray}
Similarly, we can show
\begin{eqnarray}
 \tilde l_{obs}(\phi^{(t+1)}, g^{(t)}\mid w^{*(t)})\leq\tilde l_{obs}(\phi^{(t+1)}, g^{(t+1)}\mid w^{*(t)}),\nonumber
\end{eqnarray}
using the Fr\'echet derivative.

By Monotone convergence theorem, we have 
\begin{eqnarray}
l_{obs}(\phi, g_{\phi}; \hat \eta)-\hat l_{obs}(\phi^{(t)}, g_{\phi^{(t)}}\mid w^{*(t)})\xrightarrow{}0, \nonumber\\
\tilde l_{obs}(\phi, g; \hat \eta)-\tilde l_{obs}(\phi^{(t+1)}, g^{(t)}\mid w^{*(t)})\xrightarrow{}0,\nonumber
\end{eqnarray}
in probability and for any $y$, as $t\xrightarrow{}\infty, M\xrightarrow{}\infty$.

Thus, we conclude that our proposed algorithm provides the same solutions as applying the profile likelihood method to $l_{obs}(\phi, g; \hat \eta)$ directly. Thus, our proposed estimators enjoy the same asymptotic distributions in (\ref{hat_phi}).

\section{Proof of Theorem \ref{thm:thm2}}\label{App::D}
Let $\hat \theta $ is the solution of 
\begin{eqnarray}
U(\theta\mid \hat \phi, \hat g)=\frac{1}{n}\sum_{i=1}^{n}\frac{\delta_i}{\pi\left\lbrace  x_{i1}^T\hat{\phi}+\hat g(y_i)\right\rbrace } U(\theta; x_i, y_i)=0,\label{U_theta}
\end{eqnarray} 
where $(\hat \phi, \hat g)$ is obtained from our proposed method. Note that $\hat g=\hat g_{\hat \phi}$.
Then, we apply the Taylor linearization to (\ref{U_theta}) and obtain 
\begin{eqnarray}
&U(\hat \theta\mid \hat \phi, \hat g)\cong& U(\theta_0\mid \phi_0, \hat g_0)+ \frac{\partial U(\theta_0\mid \phi_0, \hat g_0)}{\partial \theta_0}(\hat \theta-\theta_0) \nonumber\\
&&+ \frac{\partial U(\theta_0\mid \phi_0, \hat g_0)}{\partial \phi_0}(\hat \phi-\phi_0).\label{U_1}
\end{eqnarray}
Moreover, using Fr\'echet derivative, we have
\begin{eqnarray}
 U(\theta_0\mid \phi_0, \hat g_0)\cong  U(\theta_0\mid \phi_0, g_0)+\frac{\partial  U(\theta_0\mid \phi_0, g_0)}{\partial g_0}(\hat g_0-g_0).\label{U_2}
\end{eqnarray}
Using (\ref{U_1}) and (\ref{U_2}), we get the final expansion as
\begin{eqnarray}
&U(\hat \theta\mid \hat \phi, \hat g)\cong& U(\theta_0\mid \phi_0, g_0) +\frac{\partial U(\theta_0\mid \phi_0, g_0)}{\partial \theta_0}(\hat \theta-\theta_0) \nonumber\\
&& \frac{\partial U(\theta_0\mid \phi_0,  g_0)}{\partial \phi_0}(\hat \phi-\phi_0)+\frac{\partial  U(\theta_0\mid \phi_0, g_0)}{\partial g_0}(\hat g_0-g_0).\nonumber
\end{eqnarray}

From Lemma (\ref{lemmaC3}), we have 
\begin{eqnarray}
\frac{1}{\sqrt n} \left.\frac{d}{d\phi} \frac{\partial 	l_{obs}(\phi, g_{\phi}) }{\partial g}\right|_{\phi=\phi_0} (\hat g_0-g_0)=o_p(1).\nonumber
\end{eqnarray}

Assume $$\sup_{y}\left|\frac{1}{\sqrt n} \left.\frac{d}{d\phi} \frac{\partial 	l_{obs}(\phi, g_{\phi}) }{\partial g}\right|_{\phi=\phi_0}\right|=O_p(\sqrt{n}).$$
Then $\sup_{y} |(\hat g_0-g_0)|=o_p(n^{-1/2})$.

Assume $$ \sup_{y} \left|\frac{\partial U(\theta_0\mid \phi_0, g_0)}{\partial g_0}\right|=O_p(1) .$$ Then, $$ \frac{\partial  U(\theta_0\mid \phi_0, g_0)}{\partial g_0}(\hat g_0-g_0)$$ is negligible. Thus, we have 
\begin{eqnarray}
&U(\hat \theta\mid \hat \phi, \hat g)\cong& U(\theta_0\mid \phi_0, g_0) +\frac{\partial U(\theta_0\mid \phi_0, g_0)}{\partial \theta_0}(\hat \theta-\theta_0) \nonumber\\
&& \frac{\partial U(\theta_0\mid \phi_0,  g_0)}{\partial \phi_0}(\hat \phi-\phi_0),\nonumber
\end{eqnarray}
which leads to 
\begin{eqnarray}
\hat \theta-\theta_0\cong-\left[ E\left\lbrace  \frac{\partial U(\theta_0\mid \phi_0, g_0)}{\partial \theta_0} \right\rbrace \right] ^{-1} \left[  U(\theta_0\mid \phi_0, g_0) +E\left\lbrace \frac{\partial U(\theta_0\mid \phi_0,  g_0)}{\partial \phi_0}\right\rbrace (\hat \phi-\phi_0) \right] .\label{theta_hat}
\end{eqnarray}

Since 
\begin{eqnarray}
U(\theta_0\mid \phi_0,  g_0)\xrightarrow{}0\nonumber\\
\hat \phi-\phi_0\nonumber
\end{eqnarray}
in probability, we can conclude that 
\begin{eqnarray}
\hat \theta-\theta_0\xrightarrow{}0
\end{eqnarray}
in probability.

Using (\ref{theta_hat}), we have 
\begin{eqnarray}
&n \left[ E\left\lbrace  \frac{\partial U(\theta_0\mid \phi_0, g_0)}{\partial \theta_0} \right\rbrace \right] ^{-1} & \mathrm{var}\left[  U(\theta_0\mid \phi_0, g_0) +E\left\lbrace \frac{\partial U(\theta_0\mid \phi_0,  g_0)}{\partial \phi_0}\right\rbrace (\hat \phi-\phi_0) \right] \nonumber\\
&&  \times\left[ E\left\lbrace  \frac{\partial U(\theta_0\mid \phi_0, g_0)}{\partial \theta_0} \right\rbrace \right] ^{-1}\xrightarrow{}\Sigma,\nonumber
\end{eqnarray}
in probability.

Therefore, our final conclusion is that 
\begin{eqnarray}
\sqrt{n}(\hat \theta-\theta_0)\xrightarrow{}N(0, \Sigma)
\end{eqnarray}
in distribution.
\clearpage
\bibliographystyle{chicago}
\bibliography{ref}

\clearpage

\end{document}